\documentclass[fleqn,usenatbib]{mnras}

\usepackage[T1]{fontenc}
\usepackage{ae,aecompl}

\usepackage{graphicx}	% Including figure files
\usepackage{amsmath}	% Advanced maths commands
\usepackage{amssymb}	% Extra maths symbols

\newcommand{\Fig}[1]{Fig.~\ref{#1}}
\newcommand{\Sec}[1]{Section~\ref{#1}}
\newcommand{\threehun}{{\sc TheThreeHundred}}

\title[Backsplash galaxies]{The Three Hundred Project: Backsplash galaxies in simulations of clusters}

\author[R. Haggar et al.]{\parbox{\textwidth}{
Roan Haggar,$^{1}$\thanks{E-mail: roan.haggar@nottingham.ac.uk}
Meghan E. Gray,$^{1}$
Frazer R. Pearce,$^{1}$
Alexander Knebe,$^{2,3,4}$}
\newauthor{Weiguang Cui,$^{2,5}$
Robert Mostoghiu,$^{2}$
Gustavo Yepes$^{2,3}$}
\\\\
\parbox{\textwidth}{
$^{1}$School and Physics \& Astronomy, University of Nottingham, Nottingham NG7 2RD, UK\\
$^{2}$Departamento de F\'isica Te\'{o}rica, M\'{o}dulo 15 Universidad Aut\'{o}noma de Madrid, 28049 Madrid, Spain\\
$^{3}$Centro de Investigaci\'{o}n Avanzada en F\'{\i}sica Fundamental (CIAFF), Universidad Aut\'{o}noma de Madrid, 28049 Madrid, Spain \\
$^{4}$International Centre for Radio Astronomy Research, The University of Western Australia, 35 Stirling Highway, Crawley, Western Australia 6009, Australia\\
$^{5}$Institute for Astronomy, University of Edinburgh, Royal Observatory, Edinburgh EH9 3HJ, United Kingdom
}}

\date{Accepted 2020 January 24. Received 2020 January 24; in original form 2019 September 13}

\pubyear{2019}

\begin{document}
\label{firstpage}
\pagerange{\pageref{firstpage}--\pageref{lastpage}}
\maketitle

% Abstract of the paper
\begin{abstract}
In the outer regions of a galaxy cluster, galaxies may be either falling into the cluster for the first time, or have already passed through the cluster centre at some point in their past. To investigate these two distinct populations, we utilise \threehun\ project, a suite of 324 hydrodynamical resimulations of galaxy clusters. In particular, we study the `backsplash population' of galaxies; those that have passed within $R_{200}$ of the cluster centre at some time in their history, but are now outside of this radius. We find that, on average, over half of all galaxies between $R_{200}$ and $2R_{200}$ from their host at $z=0$ are backsplash galaxies, but that this fraction is dependent on the dynamical state of a cluster, as dynamically relaxed clusters have a greater backsplash fraction. We also find that this population is mostly developed at recent times ($z\lesssim0.4$), and is dependent on the recent history of a cluster. Finally, we show that the dynamical state of a given cluster, and thus the fraction of backsplash galaxies in its outskirts, can be predicted based on observational properties of the cluster. 
\end{abstract}


%%%%%%%%%%%%%%%%%%%%%%%%%%%%%%%%%%%%%%%%%%%%%%%%%%

%%%%%%%%%%%%%%%%% BODY OF PAPER %%%%%%%%%%%%%%%%%%

\section{Introduction}
\label{sec:intro}

The Lambda cold dark matter ($\Lambda$CDM) model of the Universe describes the hierarchical formation of cosmological structure. Gravitational collapse results in the formation of small, bound structures, which then grow via merger events with other haloes, and through the accretion of diffuse dark matter \citep{white1978, frenk2012}. This process results in the formation of increasingly large haloes, into which gas can fall and condense to form galaxies \citep{springel2005b}. In this paradigm, the largest gravitationally bound structures are galaxy clusters. 

It is well established that the physical properties of a galaxy are strongly dependent on the environment in which the galaxy is found. One of the first quantitative studies in this area was that of \citet{dressler1980}, who described the relative excess of early-type galaxies within cluster environments, compared to the number of late-type galaxies (which are far more common in isolated, field environments). More recent work has quantified the similar relation between star formation rate and environment \citep{thomas2010, patel2011}, and examined the evolution of these relations over time \citep{vanderwel2007}. 

A range of mechanisms exist that have the potential to explain the effects of environment on galaxies. Cosmological simulations show that ram pressure stripping is enhanced in cluster environments, and can lead to infalling galaxies being almost entirely stripped of their halo gas. This can occur even in the outskirts of a cluster, resulting in the quenching of star formation in cluster galaxies \citep{zinger2018, arthur2019}. Observational evidence for ram pressure stripping includes cluster galaxies whose molecular gas reservoirs have been disturbed, meaning the gas is distributed asymmetrically with respect to the stellar component of the galaxy \citep{zabel2019}. The most extreme examples of this, known as `jellyfish galaxies', can have long tails of stripped gas extending far beyond the visible component of the galaxy, as shown in \citet{cramer2019}. Other processes such as galaxy harassment \citep{moore1996} and starvation \citep{larson1980} are also enhanced within clusters; these are described in greater detail in \citet{boselli2006}.

However, galaxies can pass through several different environments during their lifetime, each of which can have an impact on the properties of a galaxy. For instance, galaxies falling into a cluster along a relatively high-density cosmic filament are likely to experience different environmental effects to those being accreted from the field, and galaxies that are members of an infalling group will have different histories to those that are infalling as isolated objects \citep{white2010, cybulski2014, jaffe2016}. The means by which a galaxy is affected before entering a cluster are known collectively as `pre-processing'. 

Clusters are not static objects, and do not smoothly accrete matter throughout their history. Some of the first work in which this idea was investigated was that of \citet{fillmore1984}, who numerically studied the gravitational collapse of collisionless matter haloes, and the paths that particles take on their first and subsequent infalls. More specifically, the presence of a `splashback radius' in galaxy clusters also indicates that material can leave a cluster and then re-enter at a later stage. Theoretical work on this radius includes that of \citet{adhikari2014}, \citet{diemer2014} and \citet{more2015}, who each describe the splashback radius as the distance from a cluster centre at which accreting matter first reaches the apocentre of its orbit, and show that this radius physically corresponds to the distance at which the cluster density profile drops most steeply. They then proceed to identify and study the splashback radius in simulations of clusters. Observational studies have also confirmed the presence of a splashback radius, through optical \citep{more2016, baxter2017}, weak lensing \citep{chang2018} and S-Z measurements \citep{shin2019, zurcher2019} of clusters' density profiles, although the detected radii appear to consistently take smaller values than predicted. 

Closely tied to the splashback radius are `backsplash galaxies', a population of galaxies that have fallen into a cluster, but have overshot the cluster centre and have passed back beyond a certain distance from the cluster centre. Typically, distances of either $R_{200}$ (the radius within which the mean density of a cluster is equal to 200 times the critical density of the Universe) or a definition of the virial radius of the cluster (such as that of \citet{bryan1998}) are used to define this `backsplash population'. At the present day these galaxies are outside of the cluster, either receding from the cluster centre, or on a second (or subsequent) infall. 

It is particularly important to note the distinction between the definitions for splashback and backsplash, and that these two are not interchangeable; the splashback radius is based on the radial density profile of a cluster \citep{more2015} and -- for a spherical system -- clearly separates infalling material from matter orbiting in the potential of the halo. However, a backsplash galaxy refers to an individual object that has simply left the cluster, having previously been within a given (close) distance of its centre. Consequently, the splashback radius does not necessarily include all backsplash galaxies. For example, as no assumptions are made about the boundness of backsplash galaxies, they could travel far beyond the splashback radius, rather than remaining on bound orbits. This would be analogous to the `renegade subhaloes' identified by \citet{knebe2011b} in simulations of the Local Group, which were associated with a host halo, but entered a different host at a later time.

\citet{gill2005} identify backsplash galaxies in simulations of clusters, and show that the outskirts of a cluster contain a significant population of these galaxies. These galaxies will therefore have experienced the effects of a cluster environment in their past, but are found in the same locations as infalling galaxies when observed. Because of this, samples of infalling galaxies collected by surveys of cluster outskirts are likely to be contaminated by this backsplash population, making it difficult to disentangle the effects of pre-processing and the effects of a cluster on a population of galaxies. 

Backsplash galaxies are not easy to differentiate from infalling galaxies, although the two are potentially distinguishable through kinematics \citep{gill2005, pimbblet2011}. Simulations are a vital tool for studying a backsplash population, as they allow us to examine the full history of a cluster and determine the fraction of galaxies that are indeed backsplash. In this work we use \threehun\ project, a sample of 324 resimulated galaxy clusters taken from a \mbox{$1\ h^{-1}$ Gpc} cosmological volume simulation, each with full-physics hydrodynamics. Using these simulations, we study the fraction of galaxies in the outskirts of clusters that have previously been within the cluster environment, and how this fraction depends on properties of a given cluster. We also investigate the dependence of these galaxies on the distance from the cluster centre, and the evolution of this `backsplash fraction' with redshift. Finally, we emphasise the distinction between the splashback radius, and the backsplash population which we study in this work. 

This paper is structured as follows: in \Sec{sec:methods} we present details about the cluster data used in this work, and our definition of backsplash galaxies. In \Sec{sec:results} we present our results, and discuss which cluster properties affect the population of backsplash galaxies. We then summarise our findings in \Sec{sec:conclusions}. 

\section{Simulations \& Numerical methods}
\label{sec:methods}

\subsection{Hydrodynamical simulations}
\label{sec:hydro}

The galaxy clusters making up \threehun\ dataset were generated by resimulating 324 clusters in the dark-matter-only MDPL2 MultiDark simulation \citep{klypin2016}\footnote{The MultiDark simulations are publicly available from the cosmosim database, \url{https://www.cosmosim.org}.}. The simulation consists of a box with sides of comoving length \mbox{$1\ h^{-1}$ Gpc}, and contains $3840^3$ particles each of mass \mbox{$1.5\times 10^{9}\ M_{\odot}$}. \textit{Planck} cosmology was used in the MDPL2 simulation (\mbox{$\Omega_{\rm{M}}=0.307$}, \mbox{$\Omega_{\rm{B}}=0.048$}, \mbox{$\Omega_{\Lambda}=0.693$}, \mbox{$h=0.678$}, \mbox{$\sigma_{8}=0.823$}, \mbox{$n_{\rm{s}}=0.96$}) \citep{planck2016}. 

To generate \threehun\ suite, the 324 most massive clusters at $z=0$ were chosen from MDPL2. For each cluster, the particles within a spherical region of radius \mbox{$15\ h^{-1}$ Mpc} \mbox{($\sim10R_{200}$)} at \mbox{$z=0$} were traced back to their initial positions. These dark matter particles were split into dark matter and gas particles, of masses \mbox{$m_{\rm{DM}}=1.27\times10^{9}\ h^{-1}M_{\odot}$} and \mbox{$m_{\rm{gas}}=2.36\times10^{8}\ h^{-1}M_{\odot}$} respectively, representing the dark matter and gas mass fractions. Lower-resolution particles were used beyond \mbox{$15\ h^{-1}$ Mpc}, to replicate any large-scale tidal effects on the cluster at a lower computational cost. Each cluster was then resimulated from its initial conditions using the {\sc GadgetX} code, and 129 snapshots between $z=16.98$ and $z=0$ were saved. {\sc{GadgetX}} is a modified version of the {\sc{Gadget3}} code, which is itself an updated version of the {\sc{Gadget2}} code, and uses a smoothed-particle hydrodynamics scheme to fully evolve the gas component of the simulations \citep{springel2005a, beck2016}. A more extensive, technical description of \threehun\ dataset is available in \citet{cui2018}.

\subsubsection{Tree-building}
\label{sec:tree}

The haloes and subhaloes present in each snapshot of each cluster were found using the {\sc ahf}\footnote{\url{http://popia.ft.uam.es/AHF}} halo finder \citep{gill2004, knollmann2009}, which accounts for gas, stars and dark matter, and allows properties such as luminosity and angular momentum to be generated for each halo and subhalo, as well as returning the mass contained within gas, stars and dark matter. 

The halo merger trees were built using {\sc mergertree}, a tree-builder designed as part of the {\sc ahf} package. {\sc mergertree} uses a merit function to identify haloes in successive snapshots that share particles, and then determines a main progenitor, plus other progenitors, for each halo. This tree-builder is also able to skip snapshots if a significantly more suitable main progenitor is available in an earlier snapshot, thus `patching' over gaps in the tree which would otherwise result in the halo branch being truncated. This property is particularly useful for small subhaloes passing through a larger halo, as there may be a number of snapshots in which it is difficult to identify which particles may be bound to each halo.

A limit was also placed on the change in mass permitted between snapshots, such that no halo could experience an increase in dark matter mass of more than a factor of two between successive snapshots. This was implemented to prevent events in which a subhalo close to the core of a larger halo could be mistaken for the halo core, leading to an apparent (but non-physical) increase in its mass. Further details of {\sc ahf} and {\sc mergertree} can be found in \citet{knebe2011a} and \citet{srisawat2013}.

\subsection{Backsplash population}
\label{sec:backsplash}

The definition of a `backsplash galaxy' is somewhat subjective. A common definition is one based purely on the present-day locations of galaxies; the backsplash population consists of galaxies that have passed within the virial radius of a cluster at some previous time, but are now found outside of the cluster, at some distance \mbox{$D>R_{\rm{vir}}$} from the cluster centre \citep{gill2005, bahe2013}, although the `virial radius' used in this definition is also open to interpretation. Other work uses a definition based on the dynamics of galaxies. For example, \citet{haines2015} place no radial distance constraints on their backsplash galaxies, and instead take all galaxies that have passed through the pericentre of their orbital path but have yet to reach apocentre (and hence have an outwards radial velocity) to be backsplash galaxies. However, by this definition a significant portion of their backsplash galaxies are within the virialised region of the cluster, and galaxies that have passed through the cluster centre and are on a second infall are exempt from this definition. 

We adopt a definition similar to that of \citet{gill2005}, based on the orbital history of each galaxy relative to $R_{200}$, which is the radius we use as the extent of the cluster. We categorise each galaxy in or around a cluster into one of three groups, based on their radial distance to the cluster centre at $z=0$, $D_{z=0}$, and their minimum distance to the cluster centre at any time in their history, $D_{\rm{min}}$:

\begin{itemize}
\item{$D_{z=0}=D_{\rm{min}}$ or $D_{\rm{min}}>R_{200}$: \\ 
The infalling population: Galaxies that are on their first infall towards the cluster. These are either on approximately radial paths (giving \mbox{$D_{z=0}=D_{\rm{min}}$}), or are members of infalling groups that are yet to reach $R_{200}$, which can lead to $D_{z=0}>D_{\rm{min}}>R_{200}$.\vspace{5pt}}

\item{$D_{\rm{min}}<D_{z=0}<R_{200}$: \\ 
The cluster population: Galaxies within the radius of the cluster (taken to be $R_{200}$). These are the `normal' satellite galaxies, which we consider to be members of the cluster. Many of these are on bound orbits, although galaxies that are on paths heading out of the cluster can also be included in this definition.\vspace{5pt}}

\item{$D_{z=0}>R_{200}$, $D_{\rm{min}}<R_{200}$: \\ 
The backsplash population: Galaxies that have previously fallen through the cluster, but have now exited the cluster and exist beyond $R_{200}$ at $z=0$. These can either be receding from the cluster centre, or on a subsequent infall towards the cluster centre.}
\end{itemize}

Our definition deviates slightly from that of \citet{gill2005}, who instead define backsplash galaxies relative to a larger radius, \mbox{$R_{\rm{vir}}\sim R_{100}\sim 1.4R_{200}$}. However, \citet{gill2005} also note that $90\%$ of the backsplash galaxies they identify pass within $0.5R_{\rm{vir}}$ of the cluster centre, meaning that by considering $R_{200}$, we are unlikely to neglect a large fraction of these galaxies. 

Furthermore, this definition of backsplash galaxies applies to clusters we study at $z=0$. If instead we are interested in the backsplash galaxies of a cluster observed at a redshift \mbox{$z_{\rm{obs}}>0$}, we adjust the definition by replacing $D_{z=0}$ with the radial distance from the cluster centre at $z_{\rm{obs}}$, and by replacing $D_{\rm{min}}$ with the minimum distance a galaxy has passed to the cluster centre at any redshift $z\geq z_{\rm{obs}}$. 

Specifically, we are interested in the fraction of all galaxies in the radial region $[R_{200}, 2R_{200}]$ that are members of the backsplash population, and so have previously been within $R_{200}(z)$ of the cluster centre, where $R_{200}(z)$ is the radius of a cluster at a redshift $z$. To do this, we only consider haloes and subhaloes with masses $M_{200}\geq10^{10.5}\ h^{-1}M_{\odot}$, corresponding approximately to haloes of $100$ particles within the $15\ h^{-1}$ Mpc high-resolution region around each cluster. We also place a stellar mass cut, such that we only consider haloes with $M_{\rm{star}}\geq 10^{9.5}M_{\odot}$. This is approximately equivalent to removing all galaxies with a luminosity $L<10^{8}L_{\odot}$, whilst not removing any with $L>10^{9}L_{\odot}$. Finally, we remove all objects for which \mbox{$M_{\rm{star}}\geq0.3M_{200}$}, and so contain more than $30\%$ of their mass in stars. These haloes are typically found close to the centre of larger haloes and have been highly stripped, meaning that they contain almost no dark matter. However, because these objects are so compact and are found close to the centres of other haloes, their properties (such as their radii and masses) are not well-determined by our halo finder. These unreliable objects with a high stellar mass ratio make up less than $2\%$ of all haloes within $[R_{200},2R_{200}]$, and so we make the decision to remove these objects from our analysis. By applying these three constraints to our simulations, we consider all remaining objects to be galaxies with a significant population of stars at $z=0$.

\subsection{Dynamical state}
\label{sec:ds}

\citet{cui2018} describe three parameters that are used to determine the dynamical state of each of the clusters in \threehun. These parameters are:

\begin{itemize}
\item{Centre of mass offset, $\Delta_{\rm{r}}$: The offset of the center of mass of the cluster from the density peak of the cluster halo, as a fraction of the cluster radius $R_{200}$.}
\item{Subhalo mass fraction, $f_{\rm{s}}$: Fraction of the cluster mass contained in subhaloes.}
\item{The virial ratio, $\eta$: A measure of how well a cluster obeys the virial theorem, based on its total kinetic energy, $T$, its energy from surface pressure, $E_{\rm{s}}$, and its total potential energy, $W$. It is defined as \mbox{$\eta=\left(2T-E_{\rm{s}}\right)/\left|W\right|$.}}
\end{itemize}

Further description of each of these dynamical state parameters is available in \citet{cui2018}, and more comprehensive details in \citet{cui2017}. 

\citet{cui2018} describe a cluster as being dynamically relaxed if it satisfies \mbox{$\Delta_{\rm{r}}<0.04$}, \mbox{$f_{\rm{s}}<0.1$} and \mbox{$0.85<\eta<1.15$}, and denote it as unrelaxed if it does not satisfy all of these. In order to obtain a continuous, non-binary measure of dynamical state, we combine these three parameters into a single measure of dynamical state, the so-called `relaxation' of a cluster, $\chi_{\rm{DS}}$:

\begin{equation}
    \chi_{\rm{DS}}=\sqrt{\frac{3}{\left(\frac{\Delta_{\rm{r}}}{0.04}\right)^{2}+\left(\frac{f{\rm{s}}}{0.1}\right)^{2}+\left(\frac{|1-\eta|}{0.15}\right)^{2}}}\,.
    \label{eq:relaxation}
\end{equation}

Note that for a cluster to be most relaxed, we require $\Delta_{\rm{r}}$ and $f_{\rm{s}}$ to be minimised, and $\eta\rightarrow1$ \citep{cui2017}. $\chi_{\rm{DS}}=1$ corresponds approximately to the \citet{cui2018} definition of dynamical state, such that all of the clusters they denote as `dynamically relaxed' have $\chi_{\rm{DS}}>1$.

\subsection{Subsample of clusters}
\label{sec:subsample}

We also apply two criteria to the 324 clusters whose full histories are known, to determine which (if any) of the clusters are not suitable for our analysis. The first of these is a limit on the earliest snapshot to which the main branch of the merger tree (that is, the branch describing the history of the cluster halo) can be tracked. A missing link between snapshots would result in the history of the cluster before this link being lost, resulting in any backsplash galaxies that passed through the cluster before this time being omitted from our backsplash population, and therefore artificially reducing the backsplash fraction. Of the 324, we find 17 clusters whose main branches cannot be tracked back to before $z=0.5$. These clusters are not used in our subsequent analysis. 

We also note that, due to the merit function used in building the merger trees, some halo links between snapshots are assigned incorrectly, resulting in large apparent jumps in the position of the cluster. These mismatches are particularly common in binary cluster systems or during major mergers, when two large objects of similar size are in close proximity, as the progenitors of the two haloes can easily be switched. This is problematic in our work, as although these mismatches typically only affect a small number of snapshots, a large jump in the apparent position of a cluster between snapshots results in many of the galaxies that are within $R_{200}$ of a cluster at a snapshot $n$ appearing to be outside of $R_{200}$ at a snapshot $n+1$, artificially increasing the backsplash fraction around the cluster. We find 59 clusters whose position (in box coordinates) changes by $>0.5R_{200}(z)$ between two snapshots after $z=1$, a distance we find is non-physical, given the typical time elapsing between snapshots at this redshift \mbox{($\sim0.3$ Gyr)}. 

Nine of these clusters are also members of the group whose main branches are not traceable to $z=0.5$, resulting in a total of 67 clusters that we choose to omit from our work. We judge the remaining 257 clusters to be suitable for our backsplash analysis.

Our 257 clusters have masses (dark matter and gas, including subhaloes) between \mbox{$5\times 10^{14}\ h^{-1} M_{\odot}$} and \mbox{$2.6\times 10^{15}\ h^{-1} M_{\odot}$}, and radii ($R_{200}$) between \mbox{$1.3\ h^{-1}$ Mpc} and \mbox{$2.3\ h^{-1}$ Mpc}, with median values of \mbox{$8\times 10^{14}\ h^{-1} M_{\odot}$} and \mbox{$1.5\ h^{-1}$ Mpc} respectively. Consequently, smaller clusters and groups comparable to structures such as the Local Group (of mass \mbox{$\sim10^{12}\ M_{\odot}$}) are not considered in this work \citep{penarrubia2014}.

\section{Results \& discussion}
\label{sec:results}

\begin{figure}
	\includegraphics[width=\columnwidth]{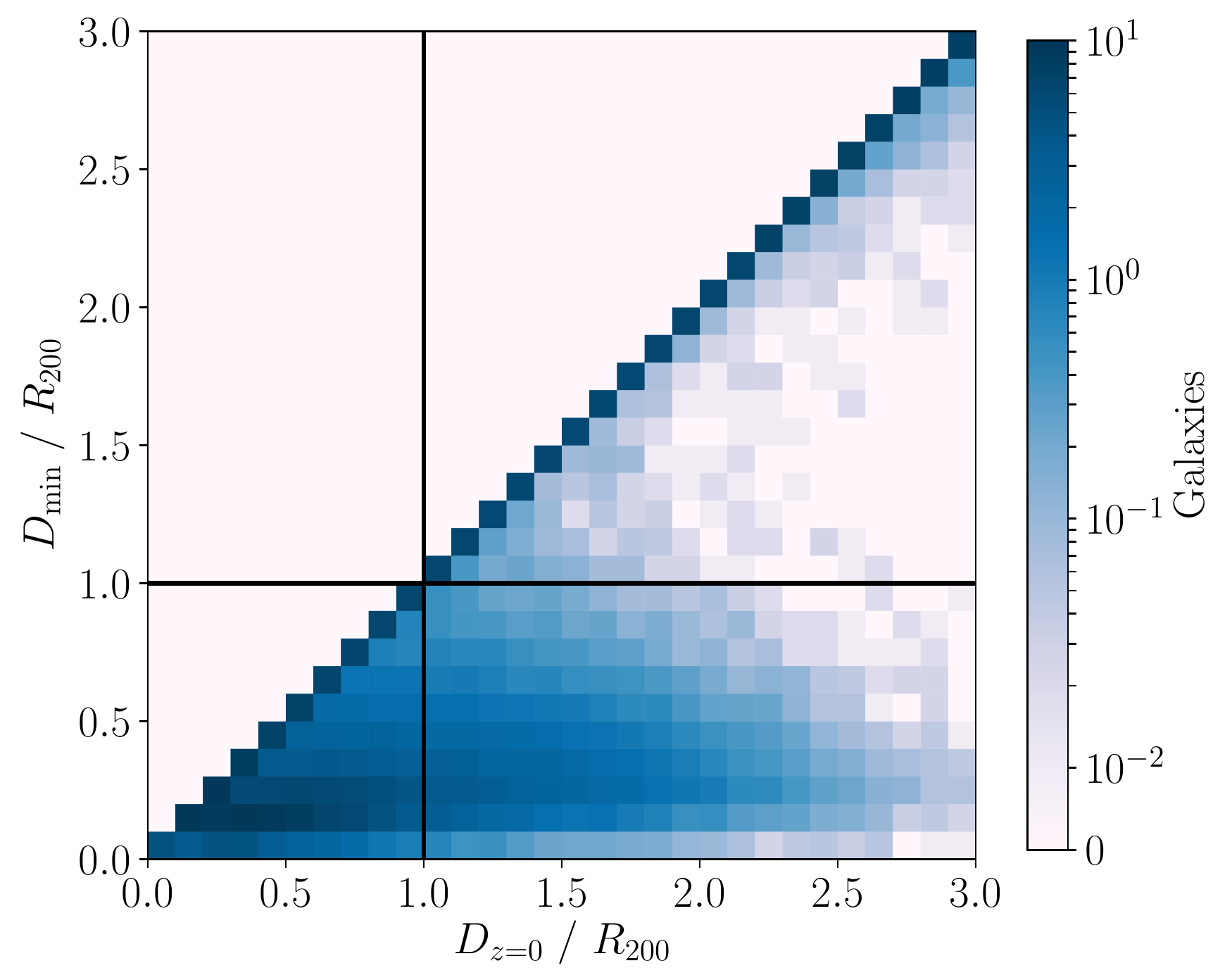}
    \caption{Phase space of galaxy population (using haloes with $M_{200}\geq 10^{10.5}M_{\odot}$, $M_{\rm{star}}\geq 10^{9.5}M_{\odot}$ and $M_{\rm{star}}<0.3M_{200}$) averaged across 257 clusters. Note the characteristic large number of objects along the line $D_{\rm{min}}=D_{z=0}$, corresponding to infalling galaxies.}
    \label{fig:backsplash257}
\end{figure}

The distribution of $D_{z=0}$ and $D_{\rm{min}}$ for each halo and subhalo in the 257 selected clusters are stacked, and are shown in \Fig{fig:backsplash257}. On average, 90\% of the backsplash haloes at $z=0$ are found between $R_{200}$ and $2R_{200}$, and 99.8\% between $R_{200}$ and $3R_{200}$. In total, we find 27114 galaxies in the radial range $[R_{200}, 2R_{200}]$, of which 15811 have previously passed within $R_{200}$ and are members of the backsplash population, corresponding to a mean backsplash fraction of 58\%. This is consistent with the work of \citet{gill2005}, who found a backsplash fraction of 50\%, albeit in a slightly different radial range, and using dark matter-only simulation data. 

\subsection{Dynamical state parameters}
\label{sec:dsres}

\begin{figure}
	\includegraphics[width=\columnwidth]{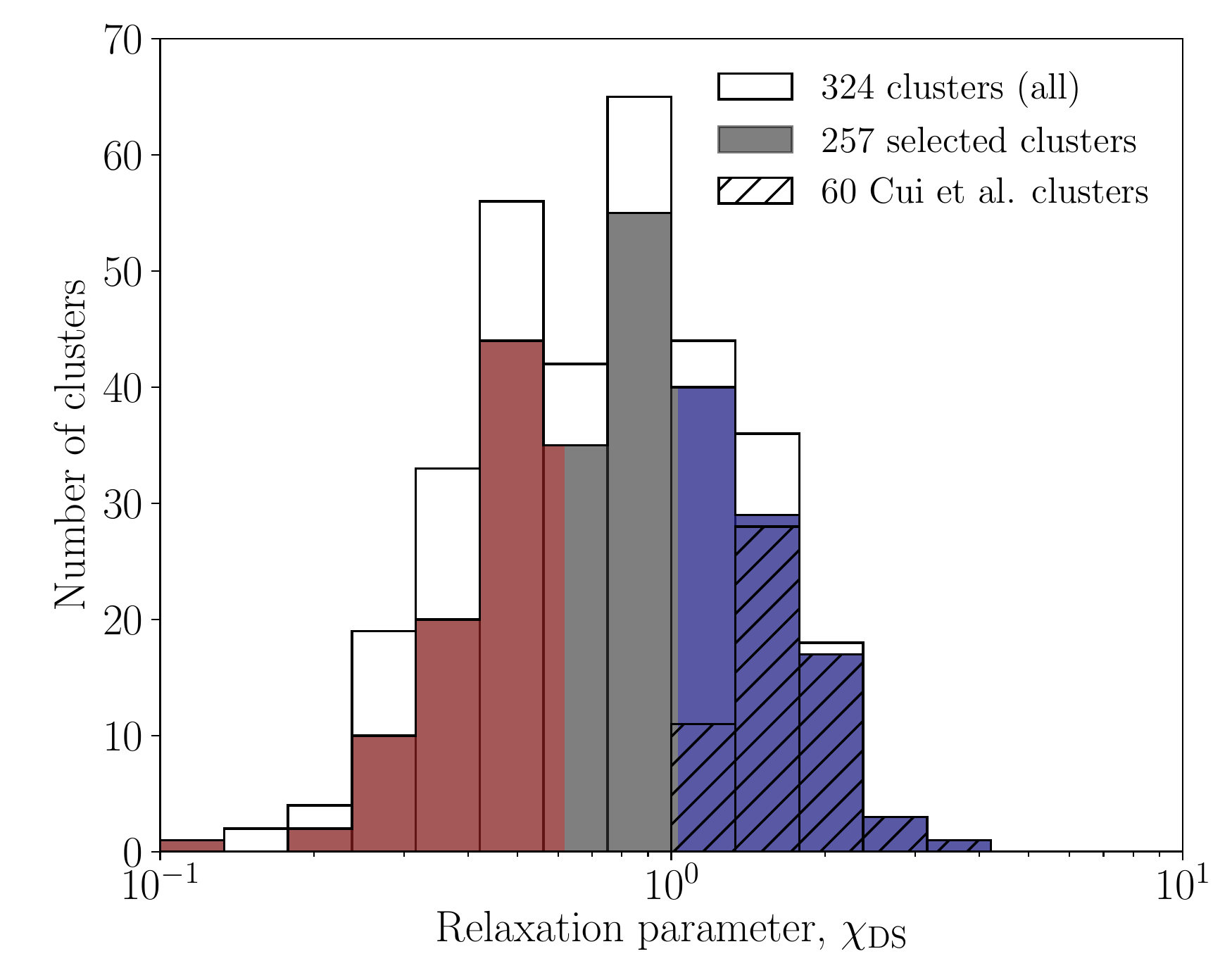}
    \caption{Distribution of the `relaxation' of each cluster, $\chi_{\rm{DS}}$, with the height of each bar showing the number of clusters in this range. The full sample of 324 clusters is shown by the white bars, and the overlaid filled bars show our selected sample of 257 clusters. The regions $\chi_{\rm{DS}}<0.619$ (`unrelaxed' clusters) and $\chi_{\rm{DS}}>1.030$ (`relaxed' clusters) are highlighted in red and blue, respectively. Our sample consists of 86 `relaxed' clusters, 86 `unrelaxed' clusters, and 85 with $0.619<\chi_{\rm{DS}}<1.030$. The hatched bars represent the clusters that are dynamically relaxed according to \citet{cui2018}; by definition, these have $\chi_{\rm{DS}}>1$.}
    \label{fig:DS_distribution}
\end{figure}

\Fig{fig:DS_distribution} shows the distribution of the dynamical state, $\chi_{\rm{DS}}$, of the 257 clusters we have selected for use in this work. The 67 clusters we remove from our original sample are slightly biased towards lower values of $\chi_{\rm{DS}}$, however they still cover most of the range of relaxation values. This is a result of the fact that the highly relaxed clusters (with greater values of $\chi_{\rm{DS}}$) are less likely to fail the selection criteria we detail in \Sec{sec:subsample}. Based on the values of $\chi_{\rm{DS}}$, we split our sample into three groups, allowing the third of clusters that are most relaxed \mbox{($\chi_{\rm{DS}}>1.030$)} and least relaxed \mbox{($\chi_{\rm{DS}}<0.619$)} to be compared. Each of these groups contains 86 clusters. The unrelaxed clusters have slightly greater average values of $M_{200}$ and $R_{200}$, however the difference is small compared to the spread of these quantities across the whole sample of clusters.

\begin{figure*}
	\includegraphics[width=\textwidth]{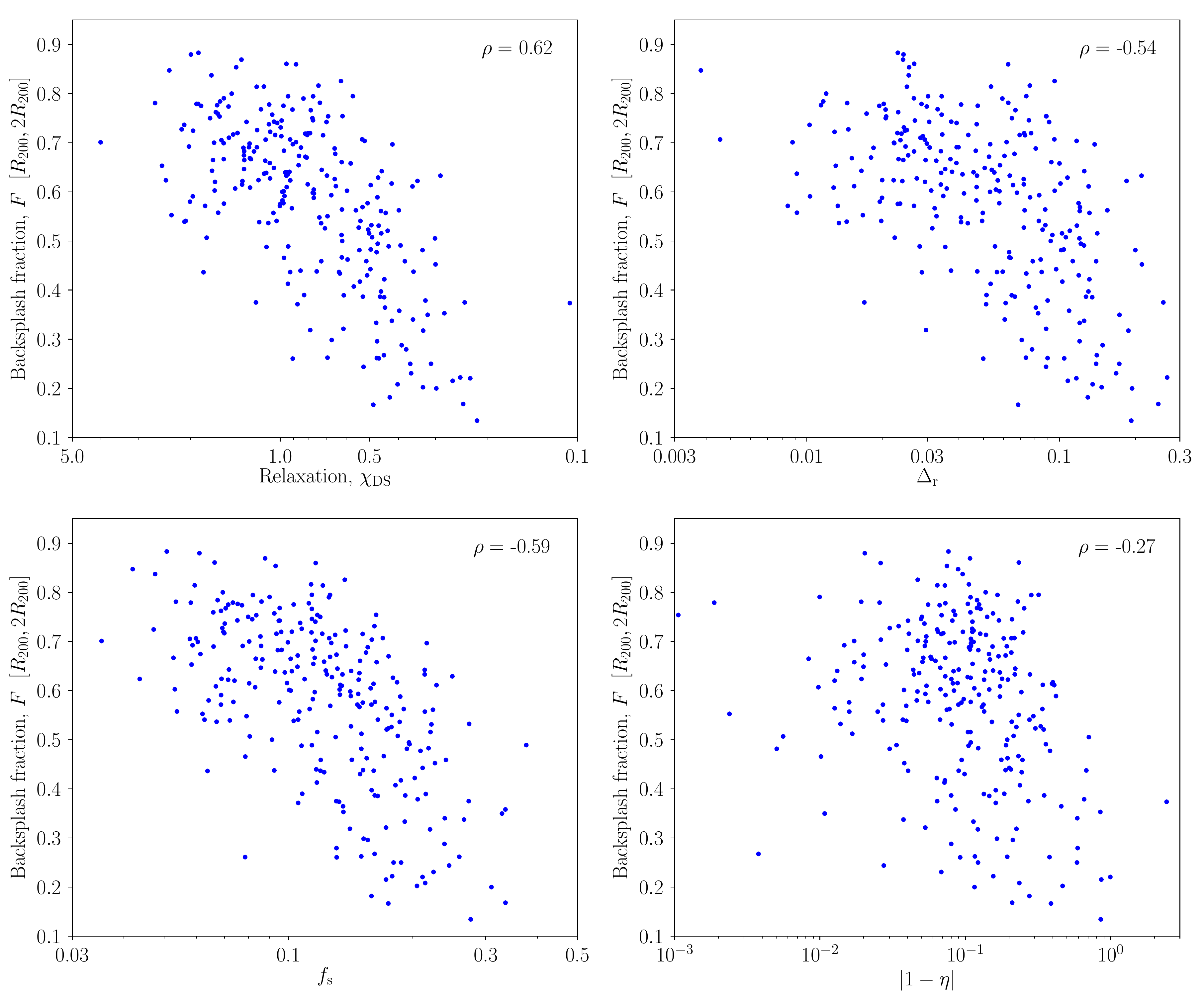}
    \caption{`Relaxation', $\chi_{\rm{DS}}$, of each of the selected 257 clusters, against backsplash fraction in the radial range $[R_{200}, 2R_{200}]$ at \mbox{$z=0$}, shown in top-left plot. The variation of backsplash fraction with each individual parameter is also shown. Note the reversed horizontal axis on the top-left plot; in each of these plots, the `more relaxed' clusters are on the left. The Spearman's rank correlation coefficient, $\rho$, for each plot is inset, showing the tighter correlation achieved by combining the three dynamical state parameters.}
    \label{fig:dynstat}
\end{figure*}

In addition to the stacked data in \Fig{fig:backsplash257}, we also calculate the backsplash fraction, $F$, for each of the 257 clusters individually. \Fig{fig:dynstat} shows the backsplash fraction for each cluster against its relaxation parameter, and shows that clusters that are more relaxed have a greater fraction of backsplash galaxies. A smaller centre of mass offset, $\Delta_{\mathrm{r}}$, smaller fraction of mass in subhaloes, $f_{\rm{s}}$, and a virial ratio, $\eta$, closer to one, all of which are indicative of a relaxed cluster, each result in a larger backsplash population.

The third of clusters that are least relaxed (with \mbox{$\chi_{\rm{DS}}<0.619$)} have a median backsplash fraction of \mbox{$F = 45^{+15}_{-20}$\%}. The most relaxed third ($\chi_{\rm{DS}}>1.030$) have a backsplash fraction $F = 69^{+9}_{-11}$\%. Of the three dynamical state parameters, $f_{\rm{s}}$ correlates most strongly with the backsplash fraction, followed by $\Delta_{\rm{r}}$. Although there is a weaker relationship between $F$ and $\left|1-\eta\right|$, a relationship does indeed exist. We note that, when considering $f_{\rm{s}}$, part of the correlation between this parameter and the backsplash fraction is caused by the backsplash galaxies themselves; the movement of a large number of galaxies from within $R_{200}$ to the cluster outskirts will reduce the amount of substructure within $R_{200}$, therefore causing the fraction of mass contained in subhaloes, $f_{\rm{s}}$, to decrease.

There is also a very weak correlation between the backsplash fraction and $M_{200}$, in which the less massive clusters have a marginally higher backsplash fraction, although this can be fully accounted for by the fact that the relaxed clusters have a slightly lower average mass. We find no significant correlation between the backsplash fraction and $R_{200}$. We therefore conclude that more dynamically relaxed clusters have greater backsplash populations, and that the fraction of mass in subhaloes and the centre of mass offset of the cluster are specific properties that we expect to affect this. 

Finally, we find that there is very little dependence of the backsplash fraction on the mass of galaxies. Separating the galaxies into stellar mass bins of width 0.5 dex, we find that the median backsplash fraction does not change for galaxies with stellar masses between \mbox{$10^{9.5}\ h^{-1}M_{\odot}$} and \mbox{$10^{11}\ h^{-1}M_{\odot}$}. Considering only galaxies with stellar masses above \mbox{$10^{11}\ h^{-1}M_{\odot}$}, the median backsplash fraction of clusters appears to drop slightly, although this drop is not statistically significant due the far smaller number of these high-mass galaxies present in cluster outskirts. However, the backsplash fraction of clusters does depend on the total masses of galaxies -- that is, the mass including the galaxy halo mass. We find that galaxies with a low halo mass are more likely to be members of the backsplash population. For example, on average $69\pm1\%$ of galaxies with total masses in the range \mbox{$[10^{10.5},10^{11}]\ h^{-1}M_{\odot}$} are backsplash galaxies, compared to $43\pm1\%$ of those with mass in the range \mbox{$[10^{12},10^{12.5}]\ h^{-1}M_{\odot}$}. Similarly, galaxies with a greater ratio of stellar mass to total mass are also more likely to be backsplash galaxies.  

We expect the halo mass and stellar mass of these galaxies to be closely linked \citep{moster2010}. As stellar material experiences very little stripping between infall and leaving a cluster in these simulations, we therefore deduce that the dark matter haloes of backsplash galaxies have been tidally stripped during their passage through the cluster. This means that the apparent bias towards low-mass haloes becoming backsplash galaxies is due to the stripping of the haloes around backsplash galaxies, rather than due to a strong dependence on the halo mass of galaxies at infall. Another potential explanation for this is dynamical friction, as previous work has shown that the location of the splashback feature in simulations of dark matter haloes is dependent on the mass of subhaloes being considered. Specifically, both \citet{adhikari2016} and \citet{more2016} find that the splashback feature for haloes of greater masses is found at smaller distances. However, for the rest of this work, we continue to focus on the effect of cluster properties on the backsplash population, rather than galaxy properties.

\subsection{Evolution of backsplash fraction}
\label{sec:evol}

By examining how the dynamical state parameters for each cluster vary over time, we are able to examine the stability of these parameters. We find that the parameters are only stable over a relatively short timescale, as the dynamical state of a cluster, given by $\chi_{\rm{DS}}$, in our $z=0$ snapshot is uncorrelated to its dynamical state in snapshots before $z=0.5$; that is, the Spearman's rank correlation coefficient between $\chi_{\rm{DS}}$ at $z=0$ and $z>0.5$ is zero. Consequently, we infer that these measures of dynamical state are dependent on the recent history of the cluster, rather than being an inherent property of the cluster that has been present since its formation. However, as \Fig{fig:dynstat} shows, the backsplash fraction is correlated with the dynamical state of a cluster. We therefore expect that the backsplash population must also be established over these relatively short timescales.

\begin{figure}
	\includegraphics[width=\columnwidth]{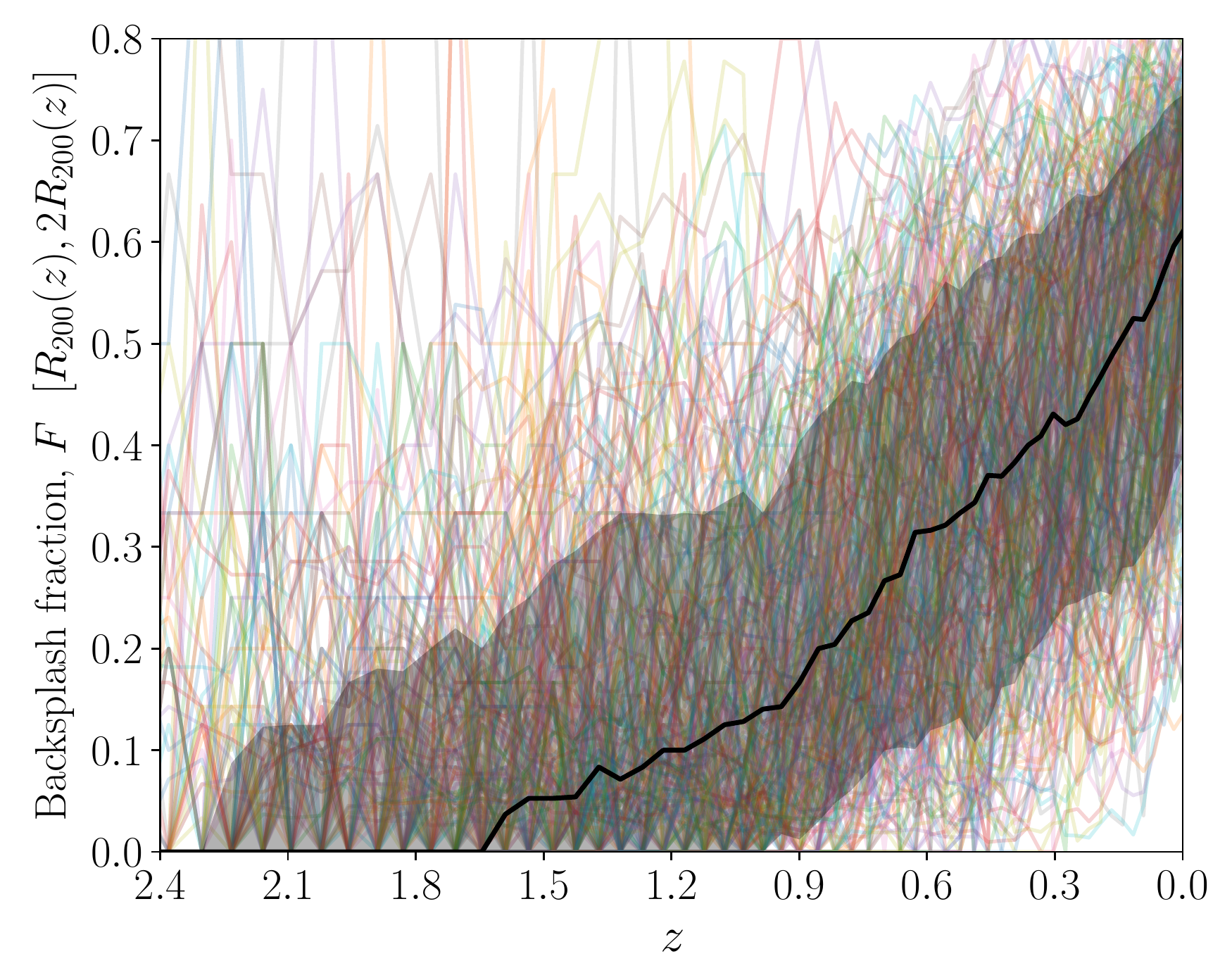}
    \caption{Evolution of backsplash fraction for 257 clusters, shown in colour. The median backsplash fraction plotted in black, and the shaded region shows the $68\%$ bounds.}
    \label{fig:bsgrowth}
\end{figure}

\Fig{fig:bsgrowth} shows that (with quite a large spread), the median backsplash fraction of each cluster is zero at $z=1.7$, and reaches half its present day value at $z=0.6$. Our definition of backsplash galaxies at $z>0$ is as we describe in \Sec{sec:backsplash}, and we define the backsplash fraction at a redshift $z_{\rm{obs}}$ as the fraction of galaxies in the radial region $[R_{200}(z_{\rm{obs}}), 2R_{200}(z_{\rm{obs}})]$ that have previously passed within $R_{200}(z)$. Consequently, if a cluster was viewed at $z>0$, this is the backsplash fraction that would be observed, based on its radius at this time. The particularly large scatter in the data at high redshifts is due mostly to our measure of backsplash fraction; if only a small ($\lesssim10$) number of galaxies are present in the outskirts of a cluster, then the presence of just one backsplash galaxy can dramatically change the value of $F$.

\begin{figure}
	\includegraphics[width=\columnwidth]{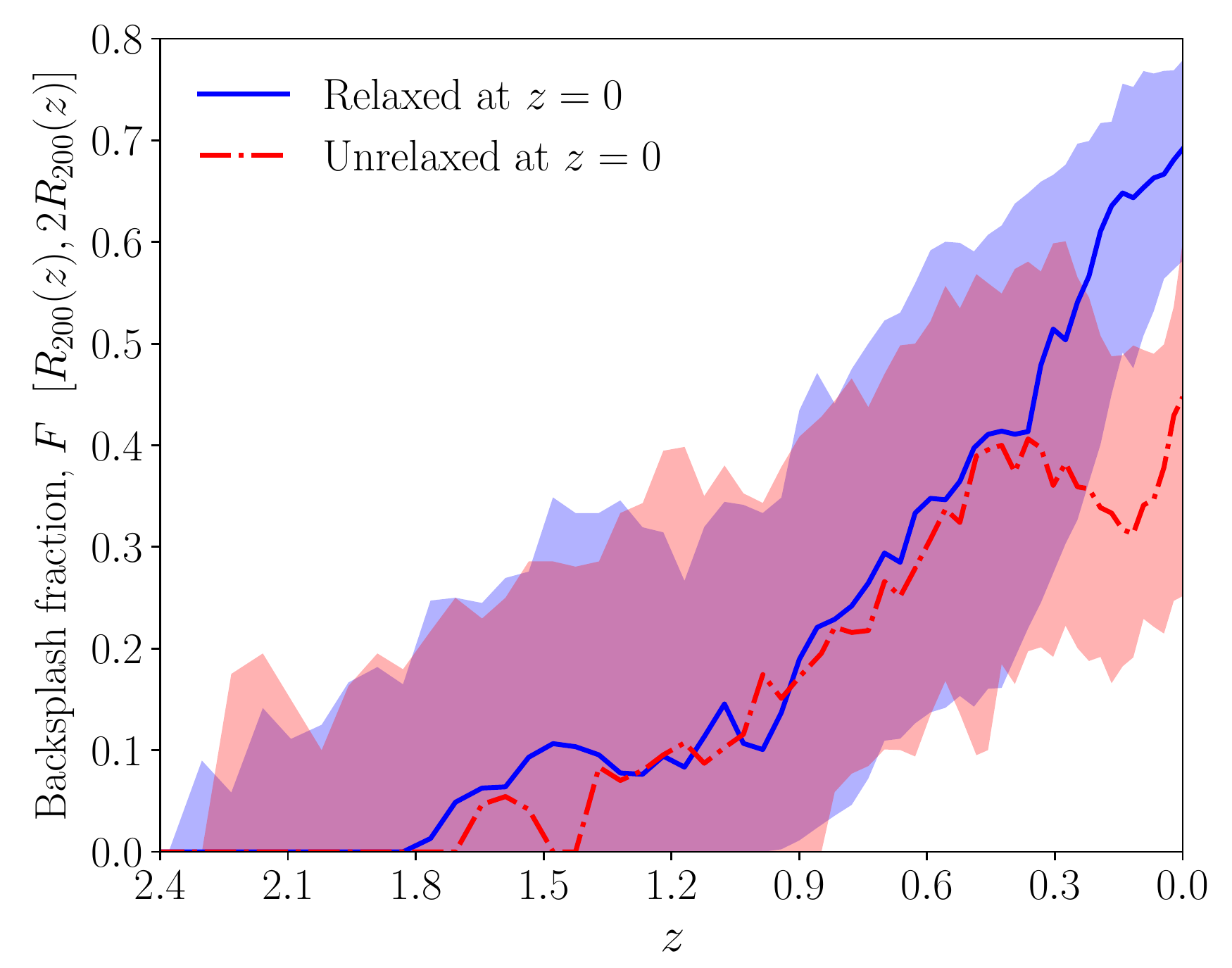}
    \caption{Median backsplash fraction over time, for 86 most relaxed ($\chi_{\rm{DS}}>1.030$) and 86 least relaxed ($\chi_{\rm{DS}}<0.619$) clusters. Note that the clusters are selected by dynamical state at $z=0$, and the same clusters are then studied at each redshift.}
    \label{fig:bsgrowth_DS}
\end{figure}

\Fig{fig:bsgrowth_DS} shows the median backsplash fraction plotted for the relaxed and unrelaxed clusters separately. Note that the clusters are selected by dynamical state at $z=0$, and the same clusters are then studied at each previous redshift. Consequently, the `relaxed' sample of clusters at $z>0$ are not necessarily those that are most relaxed at this redshift. We see that the backsplash fractions of the relaxed and unrelaxed samples agree at times before approximately $z=0.4$, when the backsplash fraction of the unrelaxed cluster sample plateaus. This indicates that the fraction is very much dependent on the recent history of a cluster, as the two types of cluster have only become distinguishable since $z=0.4$.

\begin{figure}
	\includegraphics[width=\columnwidth]{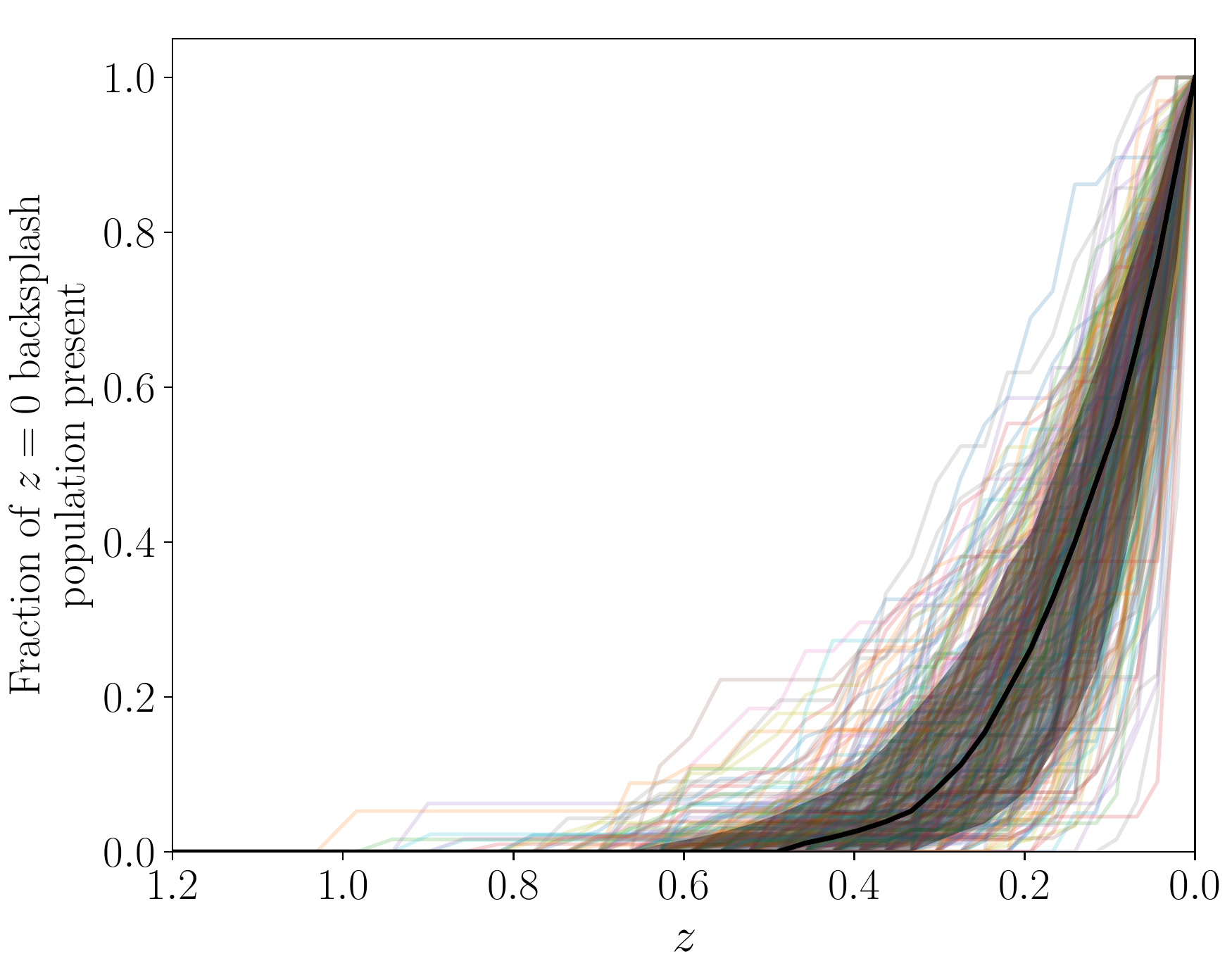}
    \caption{Fraction of backsplash population in snapshot 128 \mbox{($z=0$)} that are also members of the backsplash population at previous snapshots, for 257 clusters. Consequently, this shows the redshift at which the galaxies left the cluster, and passed outside of $R_{200}$. Median is shown in black. These times at which galaxies leave the cluster are independent of the cluster's dynamical state.}
    \label{fig:currentbs}
\end{figure}

Finally, we examine how the current backsplash population has evolved. \Fig{fig:currentbs} shows the fraction of the current backsplash galaxies that were also backsplash galaxies at previous redshifts. Note the distinction between this and \Fig{fig:bsgrowth}, as \Fig{fig:currentbs} considers only the $z=0$ backsplash galaxies, and does not include galaxies that were members of the backsplash population at $z>0$ but are within the cluster at $z=0$. For a typical cluster, the present-day backsplash all become members of the backsplash population after $z=0.5$, and half of the backsplash population is only built up at very late times ($z<0.1$). This is the case for both the relaxed and unrelaxed cluster samples. This implies that there is a dynamic population of backsplash galaxies -- a significant number of galaxies are joining and leaving the backsplash population, resulting in the overall backsplash fraction increasing relatively slowly, compared to the very rapid growth seen in \mbox{\Fig{fig:currentbs}}.

Note also that in \Fig{fig:currentbs} we consider the time since a galaxy most recently left its host cluster, meaning if a backsplash galaxy has passed through a cluster twice (and so is on its third infall), we take the time since it left the cluster for the second time (i.e. the time since it was last within $R_{200}$). Backsplash galaxies that have passed through the cluster only once make up $90\%$ of the backsplash population between $R_{200}$ and $2R_{200}$, as the typical time to cross a cluster of diameter $4$ Mpc is $\sim2$ Gyr, meaning that only backsplash galaxies that enter the cluster at very early times are able to pass through the cluster a second time. A crossing time of \mbox{2 Gyr} is consistent with \Fig{fig:currentbs}, as this period corresponds approximately to the time between $z=0.2$ and $z=0$, which as \Fig{fig:currentbs} shows, is the time over which most of the current backsplash population is built up. 

\begin{figure}
	\includegraphics[width=\columnwidth]{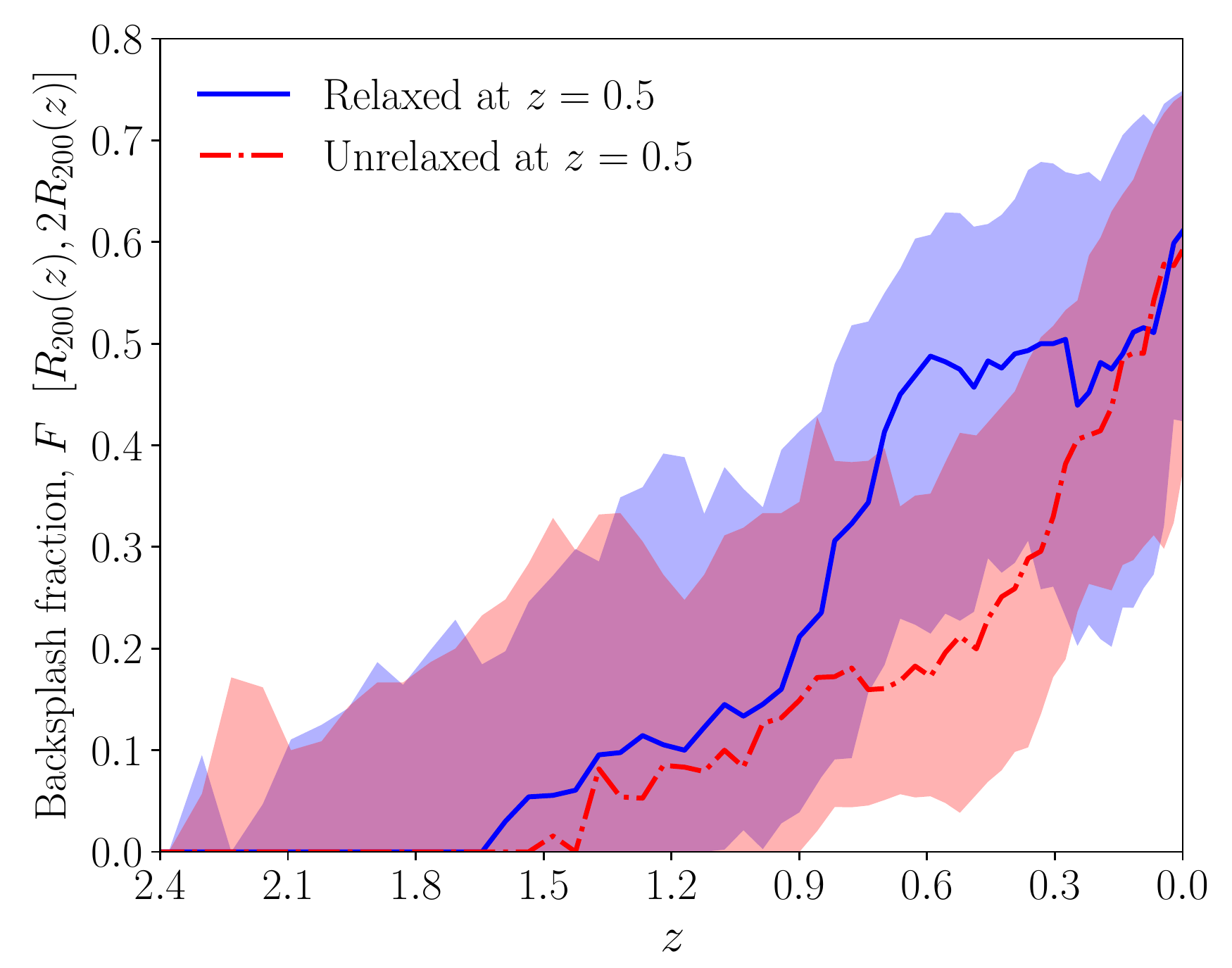}
    \caption{Median backsplash fraction over time, for clusters separated into two groups based on their values for $\chi_{\rm{DS}}$ at $z=0.5$. Backsplash fractions are for 86 most relaxed ($\chi_{\rm{DS}}(z=0.5)>0.910$) and 86 least relaxed ($\chi_{\rm{DS}}(z=0.5)<0.586$) clusters. As we make the distinction in dynamical state at $z=0.5$, the relaxed/unrelaxed samples of clusters are different to the samples used in \Fig{fig:bsgrowth_DS}.}
    \label{fig:bsgrowth_DS_z05}
\end{figure}

The recent build-up of the backsplash population also gives further support to the idea that the observed backsplash fraction of a cluster is strongly dependent on its recent history. This is corroborated by \Fig{fig:bsgrowth_DS_z05}, which shows how the backsplash fractions of clusters evolve, when the clusters are separated by dynamical state at $z=0.5$. We see comparable behaviour to \Fig{fig:bsgrowth_DS}; the backsplash populations grow at similar rates, but that of the unrelaxed clusters plateaus after $z=0.9$, whilst $F$ continues to grow for the relaxed clusters. Note that the backsplash fractions of these two cluster samples are the same at $z=0$, indicating that the dynamical state of a cluster at $z=0.5$ does not affect its backsplash population at $z=0$. 

\subsection{Role of mergers}
\label{sec:mergers}

An interpretation of the dynamical state is that it represents a measure of the formation history and growth of a cluster. For example, \citet{wen2013} determine the dynamical state of clusters based on observable quantities, and describe how large amounts of substructure in clusters (which they use as a measure of dynamical state) can be produced by major merger events. The dynamical states of our sample agree with this idea, as we find that the more relaxed clusters are those whose formation time is earlier -- that is, they are currently going through a phase of slow accretion, after an earlier phase of fast accretion. This may seem to contradict the fact that these are the clusters with a more significant backsplash population, as backsplash galaxies must have been accreted onto the cluster at recent times.

\begin{figure}
	\includegraphics[width=\columnwidth]{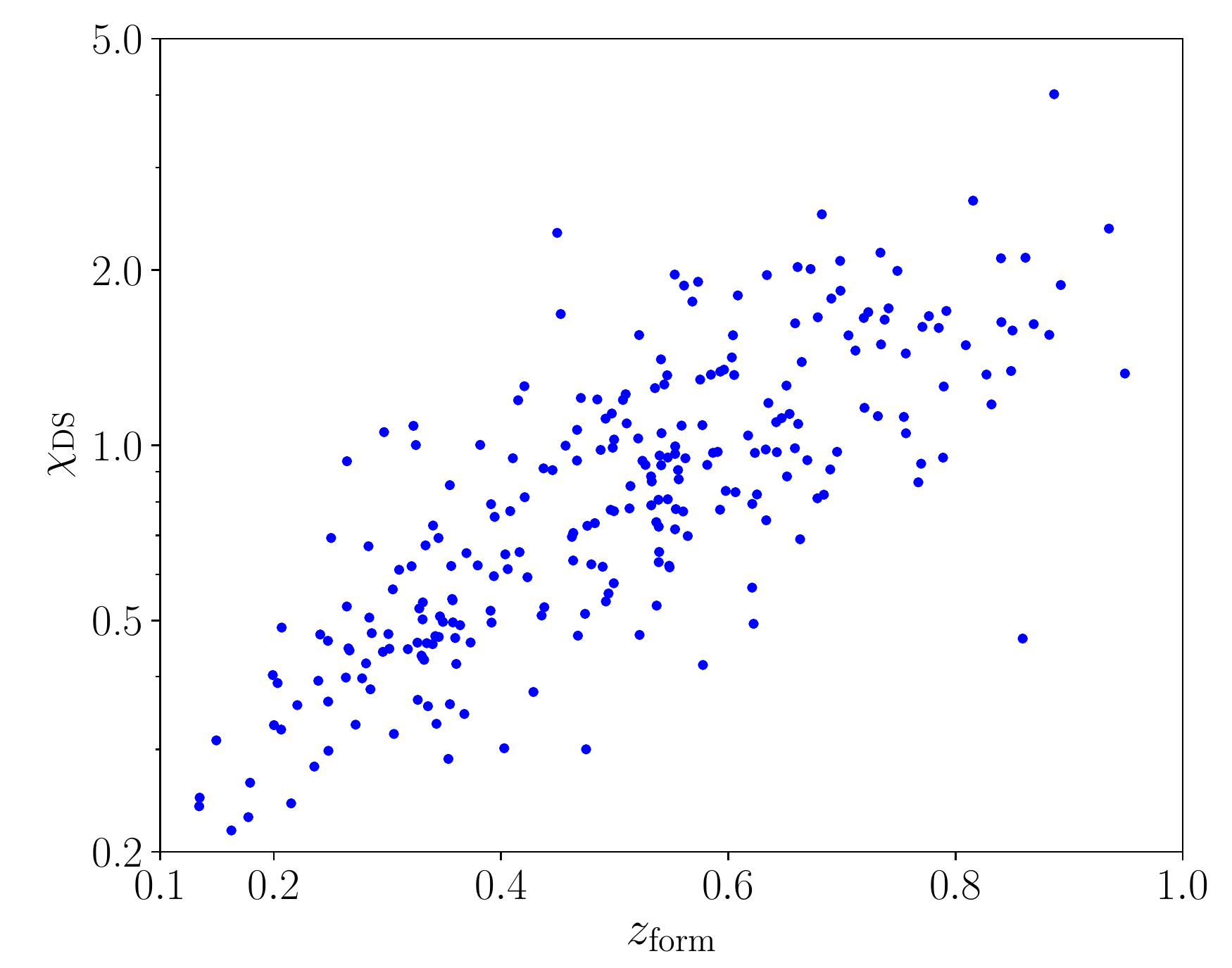}
    \caption{Relaxation at $z=0$, $\chi_{\rm{DS}}$, against formation redshift, $z_{\rm{form}}$, as defined in \citet{mostoghiu2019}, for the 257 clusters. It is clear from this that the more dynamically relaxed clusters are those that accrete much of their mass at early times ($z\gtrsim0.5$). }
    \label{fig:zform}
\end{figure}

$z_{\rm{form}}$ is the redshift at which $M_{200}$ is equal to half its value at $z=0$, as defined in \citet{mostoghiu2019}. For the relaxed sample of clusters, the average formation redshift is \mbox{$z_{\rm{form}}=0.66^{+0.17}_{-0.15}$}, and for the unrelaxed sample, \mbox{$z_{\rm{form}}=0.33^{+0.13}_{-0.09}$}. This shows that the unrelaxed sample consists of clusters that have accreted much of their mass in recent times, potentially through an event such as a major merger, and have consequently had a rapid recent growth in $M_{200}$ and $R_{200}$. \Fig{fig:zform} demonstrates that $\chi_{\rm{DS}}$ contains information about the formation history of a cluster, back to at least $z=1$. However, this does appear to contradict our previous result, that when comparing relaxed and unrelaxed clusters based on their backsplash fractions, the two types of cluster are indistinguishable before \mbox{$z\gtrsim0.4$}. 

\begin{figure}
	\includegraphics[width=\columnwidth]{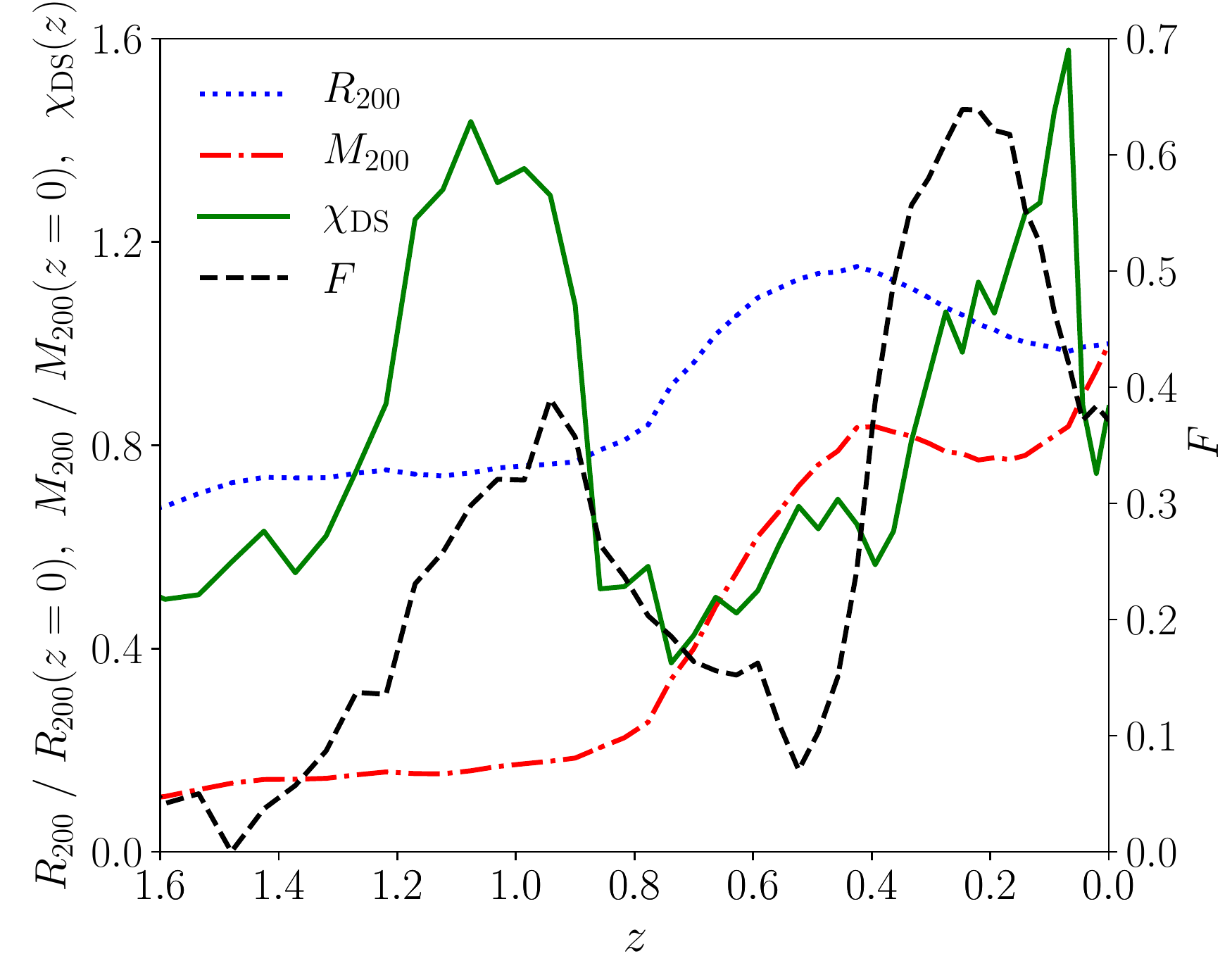}
    \caption{Evolution of the backsplash fraction, cluster radius, relaxation parameter and cluster mass (including subhalo masses) for one example cluster (with $\chi_{\rm{DS}}(z=0)=0.88$). Note the period of rapid mass accretion between $z=1.0$ and $z=0.5$, which corresponds approximately with a decrease in backsplash fraction, and with a period in which the cluster is less relaxed.}
    \label{fig:clust0007}
\end{figure}

\Fig{fig:clust0007} shows the change in $R_{200}$, $M_{200}$, $\chi_{\rm{DS}}$ and backsplash fraction, $F$, for a single cluster, as an example of this process. Between $z=1.0$ and $z=0.5$, the cluster mass increases by approximately a factor of four, and its radius increases by $50\%$, indicating a period of rapid growth. In approximately the same period, $\chi_{\rm{DS}}$ drops from a maximum value of $1.44$ (indicating a relaxed cluster) to $0.37$ (unrelaxed), and the backsplash fraction in the outskirts of this cluster drops from $39\%$ to $7\%$. This rapid increase in mass is followed by a period of near-constant cluster mass (during which the backsplash fraction and relaxation increase), and then a small merger event at $z=0.1$, which causes $F$ and $\chi_{\rm{DS}}$ to drop again.  

\begin{figure}
	\includegraphics[width=\columnwidth]{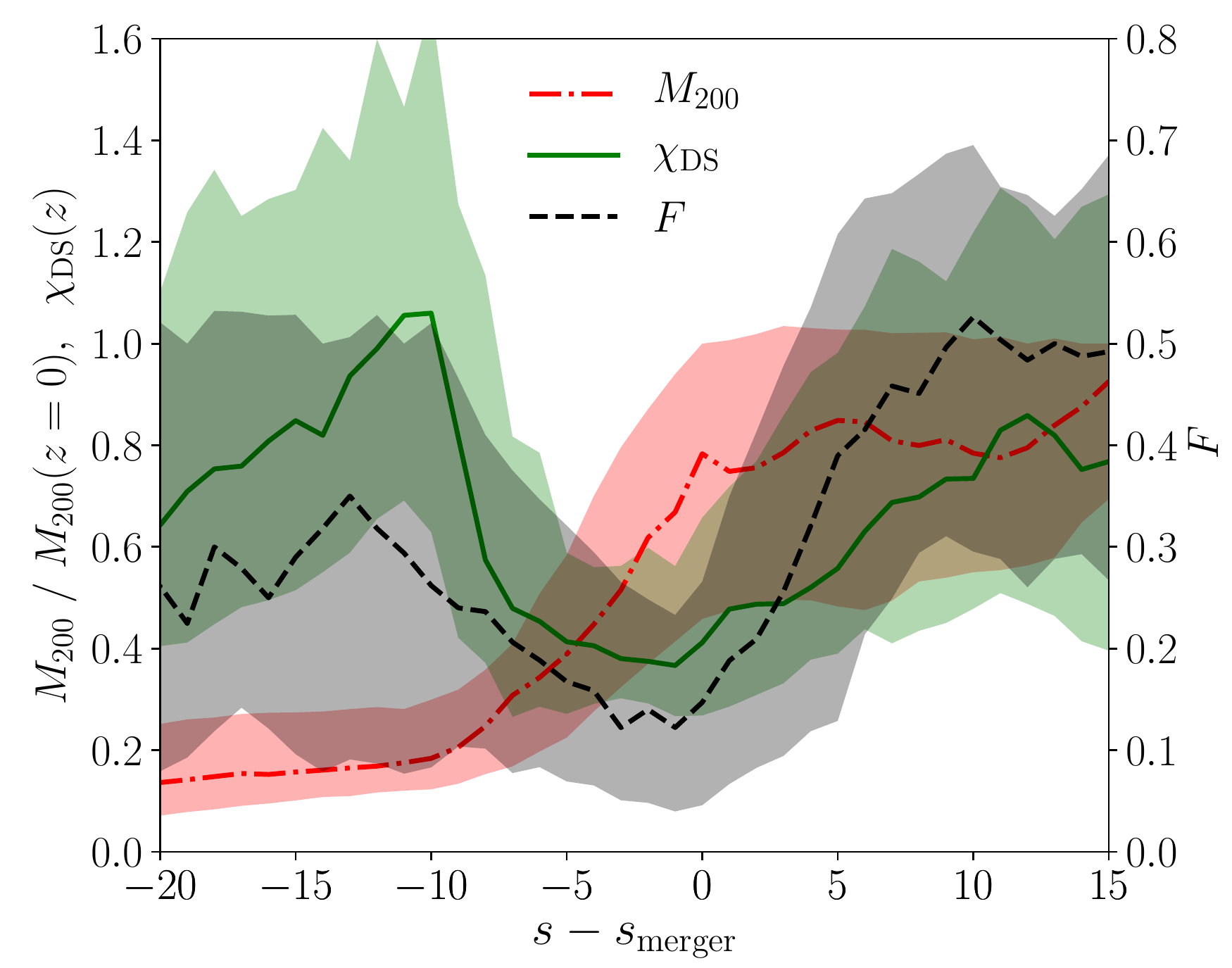}
    \caption{Stacked mass profiles, dynamical state profiles and backsplash fraction profiles, for 74 merger events in which there was a factor-of-three increase in mass within 10 snapshots at \mbox{$z<1$}. Snapshot number, $s$, is adjusted for each event relative to the snapshot at which we define the merger event to be finished, $s_{\rm{merger}}$, such that the factor-of-three mass increase occurs between $s=s_{\rm{merger}}-10$ and $s=s_{\rm{merger}}$.}
    \label{fig:merger_stack}
\end{figure}

To determine whether these periods of rapid mass accretion are responsible for the suppression of a backsplash population, we stack the mass evolution profiles and backsplash fraction histories for a large sample of clusters, as well as the evolution of their dynamical states, shown in \Fig{fig:merger_stack}. Specifically, we find 74 instances where the mass of a cluster increases by at least a factor of three within 10 snapshots ($\sim3$ Gyr) after $z=1$, and stack these 74 events. For each individual event, we select the window of 10 snapshots in which the increase in mass is greatest, and shift the snapshot numbers, $s$, such that this window corresponds to the range between $s_{\rm{merger}}-10$ and $s_{\rm{merger}}$, as demonstrated in \Fig{fig:merger_stack}. In doing so, we assume that the time elapsed between each snapshot is identical, which is approximately the case for snapshots at $z\leq1$. 

Across this selection of similar events, in 10 snapshots the median cluster mass increases by approximately a factor of four, from $0.18^{+0.12}_{-0.06}M_{200}(z=0)$ to $0.78^{+0.22}_{-0.33}M_{200}(z=0)$, either due to a series of merger events, or by a rapid period of smooth accretion. In this same period, the median backsplash fraction drops from $26^{+26}_{-18}\%$ to $15^{+12}_{-10}\%$, and the median relaxation parameter decreases from $1.06^{+0.61}_{-0.43}$ to $0.41^{+0.25}_{-0.14}$. The median cluster radius, $R_{200}$, also increases in this period, by a factor of $40\%$. This confirms that the backsplash fraction in the outskirts of a cluster is reduced during and immediately after undergoing a merger or period of rapid accretion, and that such periods also place the cluster into an unrelaxed state, although there is still a large spread in the backsplash fraction between clusters. \Fig{fig:merger_stack} also demonstrates how the reduced backsplash fraction returns within $\sim10$ snapshots to the value we would expect if a merger had not taken place, indicating that only very recent periods of rapid growth will cause the present-day backsplash population to be suppressed. 

Finally, we see from \Fig{fig:merger_stack} that the dynamical state returns to its original value over a longer timescale than the backsplash fraction ($>10$ snapshots). This difference in timescales over which $\chi_{\rm{DS}}$ and $F$ are sensitive to merger events explains the result from \Fig{fig:zform}, which shows that $\chi_{\rm{DS}}$ correlates with $z_{\rm{form}}$ for $z\leq1$, despite $F$ only being dependent on the history of the cluster for $z\leq0.4$.

\subsection{Radial backsplash profiles}
\label{sec:radial}

\begin{figure}
	\includegraphics[width=\columnwidth]{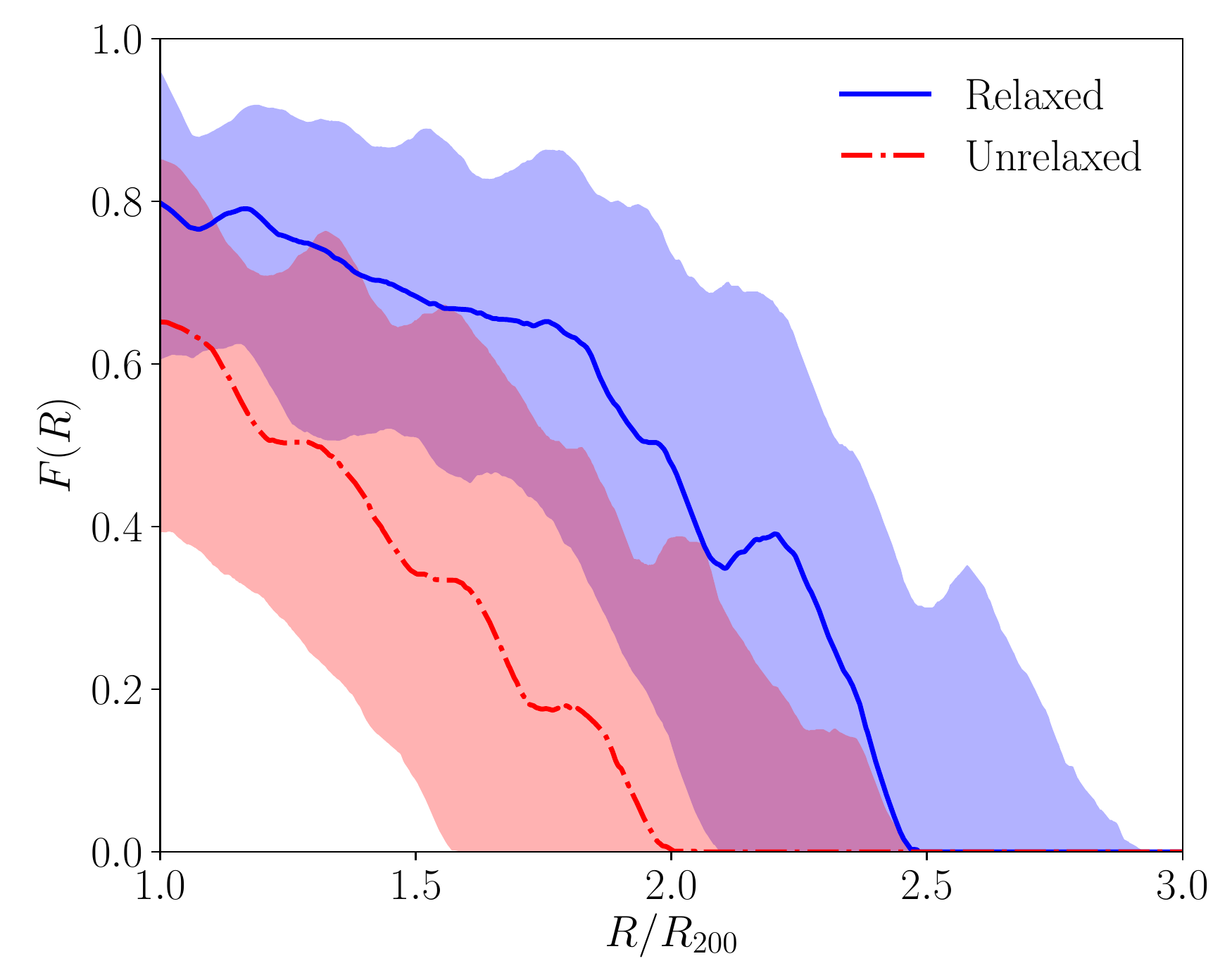}
    \caption{Median backsplash fraction as a function of radius, for relaxed and unrelaxed clusters at $z=0$. }
    \label{fig:radial_dep}
\end{figure}

To further investigate the effect of a sudden increase in cluster mass (and hence cluster radius), we also examine the radial profiles of the backsplash population. The median radial profiles of the relaxed and unrelaxed cluster samples are given in \Fig{fig:radial_dep} -- this plot shows the fraction of galaxies at a given radius that are backsplash galaxies. Note that the backsplash population is almost entirely contained within $2R_{200}$ for the unrelaxed clusters, but the relaxed clusters typically have backsplash galaxies present at distances up to $2.5R_{200}$ from the cluster centre. This is consistent with other work on the radial dependence of backsplash fractions. For example, \citet{bahe2013}, who use the same definition for backsplash fraction as us, study the radial backsplash fractions of clusters and groups using the {\sc gimic} suite of hydrodynamical simulations. Although the clusters they examine have lower masses than our sample, they describe the radial backsplash fraction of one cluster with mass $M_{200}=10^{15.2}M_{\odot}$, a typical mass for our sample of clusters. The backsplash fraction of this cluster agrees with our `relaxed' radial profile, and drops below $5\%$ at $R=2.75R_{200}$. 

This radial dependence shows why the backsplash fraction is lower in the unrelaxed clusters. As this sample have experienced a rapid increase in radius, the region we call the `cluster outskirts' ($[R_{200},2R_{200}]$) has been pushed out to greater distances, and insufficient time has passed to allow this region to be populated with backsplash galaxies. Consequently, the backsplash populations of the unrelaxed clusters are found at lower radial distances from the cluster, and fewer backsplash galaxies are present. 

Previous work has hinted at this dependence of the backsplash population on the dynamical state of a cluster, by studying the splashback radius of clusters in different dynamical states, and accreting material at different rates. Numerous studies of the splashback feature in $N$-body simulations \citep{diemer2014, diemer2017b} and in models of collapsing dark matter haloes \citep{adhikari2014, more2015} have found that the ratio between the splashback radius, $R_{\rm{sp}}$, and $R_{200}$ is smaller in rapidly-accreting, unrelaxed clusters. \citet{diemer2014} go on to explain that these clusters also typically form at later times, as indicated by \Fig{fig:zform}. This reduction in $R_{\rm{sp}}$ relative to $R_{200}$ appears to be analogous to our findings, that the backsplash population in unrelaxed clusters does not extend as far from the cluster centre as it does in relaxed clusters, resulting in a lower backsplash fraction around these clusters. 

Generally, it appears that for a particularly large backsplash fraction to build up, a cluster must remain in a relaxed, stable state for extended period of time, to allow a significant amount of infalling galaxies to pass through and join the backsplash population.

\subsection{Observational analogues for backsplash}
\label{sec:observational}

As shown in \Fig{fig:dynstat}, along with the general relaxation parameter $\chi_{\rm{DS}}$, the fraction of mass in subhaloes, $f_{\rm{s}}$, also correlates with backsplash fraction; relaxed, early-forming clusters with high backsplash fractions have a lower fraction of mass in subhaloes ($f_{\rm{s}}=0.08^{+0.03}_{-0.02}$, compared to $0.19^{+0.05}_{-0.04}$ for the unrelaxed sample). This is consistent with the work of \citet{wu2013}, who show that clusters with earlier formation times have a lower fraction of their mass stored in subhaloes. 

\begin{figure*}
	\includegraphics[width=\textwidth]{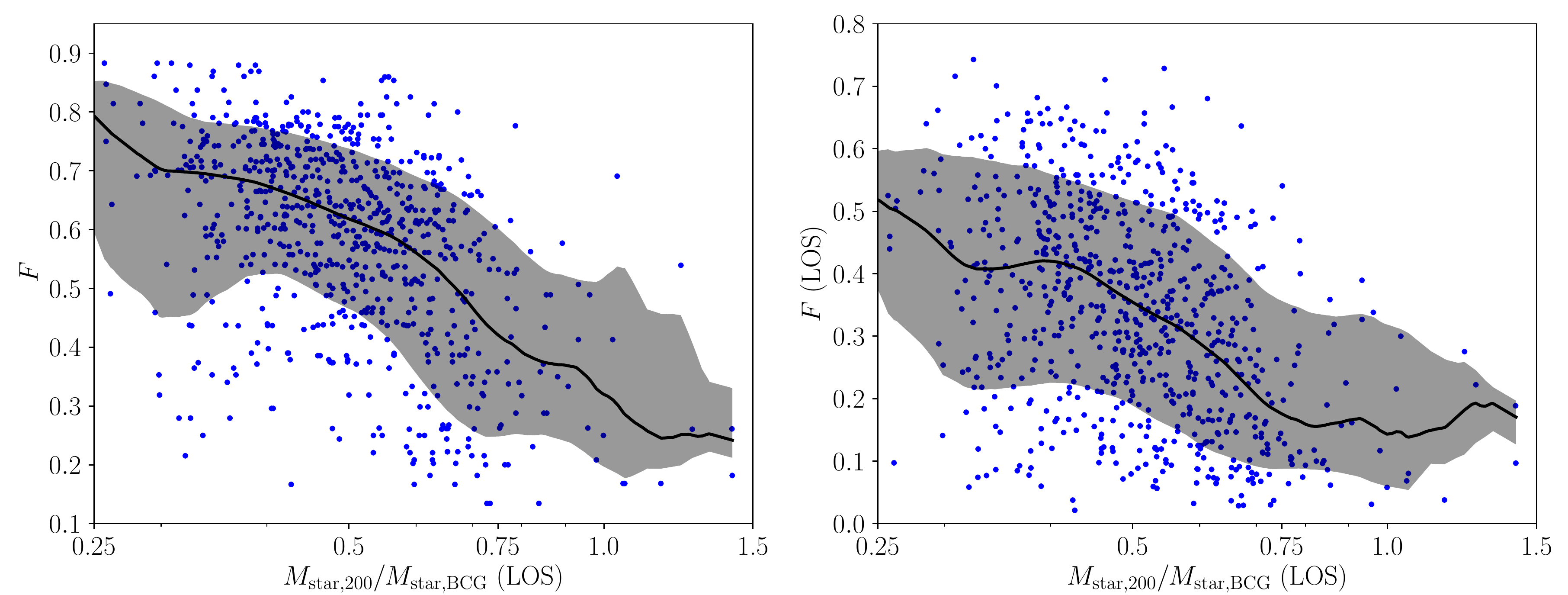}
    \caption{Ratio of stellar mass between $0.2R_{200}$ and $R_{200}$, $M_{\rm{star,200}}$, to stellar mass within $0.2R_{200}$, $M_{\rm{star,BCG}}$, against backsplash fraction, $F$, for 257 clusters. Left panel shows values of $F$ calculated using full 3D data on each cluster, and hence is the backsplash fraction in the radial region $[R_{200}, 2R_{200}]$. Right panel shows backsplash fractions for a line-of-sight (LOS) projection; that is, the fraction of galaxies in a 2D projected annulus between $[R_{200}, 2R_{200}]$ that have previously passed within $R_{200}$ in 3D space. In both plots, $M_{\rm{star,200}}$ and $M_{\rm{star,BCG}}$ are found from line-of-sight projections at $z=0$. Three orthogonal lines-of-sight are used for each cluster, which have different LOS stellar mass ratios and LOS backsplash fractions, but the same intrinsic 3D backsplash fraction. Consequently, 771 data points are used in each panel. The median backsplash fraction as a function of the stellar mass ratio is also shown for each plot.}
    \label{fig:mstar}
\end{figure*}

Directly measuring the total fraction of mass within subhaloes of a cluster with good accuracy is non-trivial. However, the luminous material within galaxies can be detected, and because \threehun\ clusters use full-physics hydrodynamics, we are able to consider what properties of the clusters can be determined, based on observable quantities. Specifically, we are interested in which measures of the dynamical state can be determined. As the distribution of satellite galaxies in a cluster is a result of the distribution of subhaloes, we use the total stellar mass of galaxies, which is detectable by cluster surveys, as a proxy for the fraction of mass in subhaloes. 

We make a radial cut at $3R_{200}$ around each cluster, and project the galaxy positions along a line-of-sight (LOS). We then take the total stellar mass of all galaxies found in the radial region $[0.2R_{200}, R_{200}]$, given by $M_{\rm{star,200}}$, and divide this by the total stellar mass within $0.2R_{200}$ of the cluster centre, $M_{\rm{star,BCG}}$. Note that we keep the same constraints on galaxies as are used throughout this work, such that we only consider galaxies with $M_{\rm{star}}\geq 10^{9.5}M_{\odot}$. The total stellar mass within $0.2R_{200}$ corresponds approximately to that of the brightest cluster galaxy (BCG), which is usually within $0.1R_{200}$ of the cluster centre, and whose brightness provides a measure of the total cluster mass \citep{lin2004}. Using this ratio as a measure of the fraction of total mass in subhaloes (and hence, as a measure of dynamical state) is in line with the work of \citet{wen2013}, who use the steepness of the cluster's radial brightness profile as a measure of dynamical state, and describe how the light of relaxed clusters tends to be dominated by the stellar material of a single, very luminous BCG.

We find that for all of the clusters except for three, $M_{\rm{star,200}}$ is between $0.25M_{\rm{star.BCG}}$ and $1.5M_{\rm{star,BCG}}$. \Fig{fig:mstar} shows the variation of backsplash fraction with this ratio of stellar masses, both in absolute terms, and with the $z=0$ galaxy positions projected along the line-of-sight onto an observational plane, such that the line-of-sight backsplash fraction represents the fraction of galaxies in an observed annulus between $R_{200}$ and $2R_{200}$ that are backsplash galaxies. 

We see that clusters with a low stellar mass between $0.2R_{200}$ and $R_{200}$, relative to the stellar mass within their inner region, have a greater backsplash fraction than clusters with large populations of satellite galaxies containing large amounts of stellar material. For example, for clusters with $M_{\rm{star,200}}=0.5M_{\rm{star,BCG}}$, the median backsplash fraction is $62^{+12}_{-15}\%$, or $35^{+16}_{-17}\%$ as measured along the line-of-sight. However, for clusters with $M_{\rm{star,200}}=M_{\rm{star,BCG}}$, we find that $F=32^{+20}_{-12}\%$, equivalent to $14^{+17}_{-8}\%$ along the line-of-sight. This agrees with the trend observed in \Fig{fig:dynstat}, where clusters with less mass contained in subhaloes (and hence more mass in the central cluster region, compared to in satellite galaxies) have greater populations of backsplash galaxies. The lower line-of-sight backsplash fractions are also as expected, due to the presence of greater-distance interlopers which are less likely to be members of the backsplash population. 

We therefore conclude that this observable quantity, $M_{\rm{star,200}}/M_{\rm{star,BCG}}$, can act as a proxy for both the intrinsic backsplash fraction of a cluster, and for the backsplash contamination of its observed outskirts, indicating a method that cluster surveys would be able to use in order to account for the backsplash fraction when studying clusters. However, it is important to note that, due to the large spread in these data, this method would be best applied to large ensemble studies of many clusters, rather than to explain trends observed within individual clusters. 

Clearly, in calculating these backsplash fractions, we assume that the radius of each cluster, $R_{200}$, is known exactly. However, the ratio of stellar masses is not strongly dependent on the value of $R_{200}$ which is used. All of the clusters have radii between $1.3\ h^{-1}$ Mpc and $2.3\ h^{-1}$ Mpc, with a median $R_{200}$ of $1.5\ h^{-1}$ Mpc. If, instead, we assume this median value as the radius of each cluster, then the impact on the observational ratios shown in \Fig{fig:mstar} is relatively small; 95\% of the clusters experience a $<20\%$ change in the ratio between $M_{\rm{star,200}}$ and $M_{\rm{star,BCG}}$, corresponding to a negligible difference in the inferred backsplash fraction, and further demonstrating that this stellar mass ratio is a potential means for estimating the backsplash fraction of clusters.

\section{Conclusions}
\label{sec:conclusions}

In this work, we have determined the fraction of galaxies in the outskirts of a cluster that are members of the `backsplash population': galaxies that have travelled along a path through the centre of a cluster and now reside in its outskirts, either receding from the cluster centre or on a subsequent infall. We have studied the time-dependence and radial-dependence of this population, discussed physical processes that could impact the prevalence of these objects based on our definition of dynamical state, and proposed observable quantities that could reveal the backsplash population of clusters in future surveys. Our findings are summarised below: 

\begin{itemize}

\item{Across the 257 clusters we consider, $58\%$ of galaxies in the radial region $[R_{200}, 2R_{200}]$ around a cluster are backsplash galaxies. However, there is a large variation in this fraction between clusters; $95\%$ of the clusters have a backsplash fraction between $21\%$ and $85\%$.}

\item{Clusters that are dynamically relaxed have a higher fraction of backsplash galaxies. Across our sample, approximately $70\%$ of galaxies (between $R_{200}$ and $2R_{200}$) in the third of clusters we deem `most relaxed' are backsplash, compared to $45\%$ in the `least relaxed' third of clusters. For the least relaxed decile, this fraction drops even further, to a median value of approximately $30\%$.}

\item{50\% of the backsplash galaxies at the present time only become backsplash galaxies (i.e. leave the region \mbox{$R<R_{200}$}) after $z=0.1$, and less then $10\%$ of the current backsplash galaxies have been members of the backsplash population since before $z=0.3$. Consequently, the $z=0$ backsplash population is strongly dependent on the recent history of a cluster. In particular, clusters that have experienced a large increase in mass (and hence $R_{200}$) at recent times, either through rapid accretion or a major merger, have a suppressed backsplash fraction. Typically, we find that a cluster increasing in mass by a factor of three over $\sim 3$ Gyr will experience greater than a factor of two drop in backsplash fraction over the same period. We find that the clusters with a large increase in mass at late times (that is, a lower formation redshift, $z_{\rm{form}}$) are less relaxed, and so contain a lower backsplash fraction.}

\item{The backsplash galaxies in clusters we identify as `unrelaxed' are mostly found within $2R_{200}$, whilst the backsplash population in `relaxed' clusters extends to distances of $2.5R_{200}$. Almost no backsplash galaxies exist beyond $3R_{200}$.}

\item{Measuring the stellar mass in a galaxy cluster relative to the stellar mass in its central region can allow the backsplash fraction of a cluster to be estimated, with an absolute uncertainty of approximately $10\%$. Clusters with outer regions containing large amounts of stellar material typically have a backsplash fraction of $30\%$, which rises to $60\%$ for clusters dominated by a bright BCG. The backsplash fraction as measured by an observer -- the fraction of galaxies in a projected annulus that are members of the backsplash population -- can also be inferred by this measure, although not with the same precision. Typically, this line-of-sight backsplash fraction is approximately a factor of two less than the absolute backsplash fraction.}

\end{itemize}

Our findings demonstrate that backsplash galaxies are likely to have a significant impact when studying the history of galaxies in cluster environments, and make it challenging to disentangle the effects of pre-processing and of a cluster environment. It should be noted that this work does not address whether a given galaxy can be identified as a backsplash galaxy or an infalling galaxy, however as properties such as galaxy luminosities in various bands are available for \threehun\ simulations (which have also been run using various other physics models, in addition to {\sc GadgetX}), we will be able to investigate this in future work. Nevertheless, we show that the backsplash population can be accounted for within cluster surveys, allowing the contamination of an infalling sample of galaxies by backsplash galaxies to be quantified, and thus allowing corrections to be made to radial profiles of galaxy properties in cluster environments.

Throughout this work, we are primarily interested in backsplash galaxies. We emphasise that it is important to distinguish between these backsplash galaxies, and the splashback radius which has been discussed in other recent work \citep{more2015}. The splashback radius is typically used to refer to an outer radius of a cluster, beyond which material is not expected to be virialised. The backsplash population instead refers to individual galaxies that have passed through the cluster, which can be difficult to identify observationally, but can be studied using simulations. As shown in this work, backsplash galaxies can be found at far greater distances \mbox{($\sim 2.5R_{200}$)} than the typical splashback radius of a cluster \mbox{($\sim 1.5R_{200}$)}.

This indicates that $R_{\rm{sp}}$ does not necessarily represent a hard boundary, outside of which all objects are infalling, in the same way that $R_{200}$ does not represent such a boundary. This is in agreement with the work of \citet{diemer2017a}, who shows that the splashback radius typically contains the apocentre of approximately 87\% of particles within a dark matter halo, and so some particles can indeed pass beyond it. However, by considering bound haloes rather than single particles, and by including hydrodynamical effects, we have shown that a significant fraction of galaxies are also expected to pass through the cluster and travel back out to relatively distant regions.

Other potential contributions to the apparent discrepancy between splashback radius and backsplash galaxies include different typical infall speeds of unbound gas and bound galaxy haloes \citep{bahe2013}, which could also be indicated by the presence of an accretion shock around clusters at a similar distance to the splashback radius \citep{arthur2019}. Anisotropy in the backsplash population could also explain how backsplash galaxies can be found beyond the splashback radius; a dependence of backsplash galaxies on large-scale structure around clusters could result in these galaxies being present at greater distances in different angular regions of the cluster outskirts, and therefore mean that a spherical splashback radius is still an insufficient method to determine the region of a cluster's influence. We will investigate anisotropy in the backsplash population in future work, as well as how gravitationally bound these types of objects are to their host clusters.

\section*{Acknowledgements}

We thank the anonymous referee for their useful comments and suggestions, which have helped to improve the quality and clarity of this paper. 

This work has been made possible by \threehun\ collaboration\footnote{\url{https://www.the300-project.org}}. This work has received financial support from the European Union's Horizon 2020 Research and Innovation programme under the Marie Sk\l{}odowskaw-Curie grant agreement number 734374, i.e. the LACEGAL project\footnote{\url{https://cordis.europa.eu/project/rcn/207630\_en.html}}. \threehun\ simulations used in this paper have been performed in the MareNostrum Supercomputer at the Barcelona Supercomputing Center, thanks to CPU time granted by the Red Espa\~nola de Supercomputaci\'on. 

RH acknowledges support from STFC through a studentship. AK, RM and GY are supported by the \textit{Ministerio de Econom\'ia y Competitividad} and the \textit{Fondo Europeo de Desarrollo Regional} (MINECO/FEDER, UE) in Spain through grant AYA2015-63810-P. AK and GY would also like to thank MINECO/FEDER (Spain) for financial support under research grants AYA2015-63819-P and PGC2018-094975-C2. AK further acknowledges support from the Spanish Red Consolider MultiDark FPA2017-90566-REDC and thanks East Village for precious diamond tears.

The authors contributed to this paper in the following ways: RH, MEG, FRP and AK formed the core team. RH analysed the data, produced the plots and wrote the paper along with MEG and FRP. AK produced the halo merger trees. WC produced the dynamical state parameters. RM provided the cluster formation times used in \Fig{fig:zform}. GY supplied the simulation data. All authors had the opportunity to comment on the paper.

%%%%%%%%%%%%%%%%%%%%%%%%%%%%%%%%%%%%%%%%%%%%%%%%%%

%%%%%%%%%%%%%%%%%%%% REFERENCES %%%%%%%%%%%%%%%%%%

\bibliographystyle{mnras}
\bibliography{backsplash} % if your bibtex file is called example.bib

\begin{thebibliography}{}
\makeatletter
\relax
\def\mn@urlcharsother{\let\do\@makeother \do\$\do\&\do\#\do\^\do\_\do\%\do\~}
\def\mn@doi{\begingroup\mn@urlcharsother \@ifnextchar [ {\mn@doi@}
  {\mn@doi@[]}}
\def\mn@doi@[#1]#2{\def\@tempa{#1}\ifx\@tempa\@empty \href
  {http://dx.doi.org/#2} {doi:#2}\else \href {http://dx.doi.org/#2} {#1}\fi
  \endgroup}
\def\mn@eprint#1#2{\mn@eprint@#1:#2::\@nil}
\def\mn@eprint@arXiv#1{\href {http://arxiv.org/abs/#1} {{\tt arXiv:#1}}}
\def\mn@eprint@dblp#1{\href {http://dblp.uni-trier.de/rec/bibtex/#1.xml}
  {dblp:#1}}
\def\mn@eprint@#1:#2:#3:#4\@nil{\def\@tempa {#1}\def\@tempb {#2}\def\@tempc
  {#3}\ifx \@tempc \@empty \let \@tempc \@tempb \let \@tempb \@tempa \fi \ifx
  \@tempb \@empty \def\@tempb {arXiv}\fi \@ifundefined
  {mn@eprint@\@tempb}{\@tempb:\@tempc}{\expandafter \expandafter \csname
  mn@eprint@\@tempb\endcsname \expandafter{\@tempc}}}

\bibitem[\protect\citeauthoryear{{Adhikari}, {Dalal}  \&
  {Chamberlain}}{{Adhikari} et~al.}{2014}]{adhikari2014}
{Adhikari} S.,  {Dalal} N.,   {Chamberlain} R.~T.,  2014, \mn@doi [\jcap]
  {10.1088/1475-7516/2014/11/019}, 11, 019

\bibitem[\protect\citeauthoryear{Adhikari, Dalal  \& Clampitt}{Adhikari
  et~al.}{2016}]{adhikari2016}
Adhikari S.,  Dalal N.,   Clampitt J.,  2016, \mn@doi [\jcap]
  {10.1088/1475-7516/2016/07/022}, 2016, 022

\bibitem[\protect\citeauthoryear{{Arthur} et~al.,}{{Arthur}
  et~al.}{2019}]{arthur2019}
{Arthur} J.,  et~al., 2019, \mn@doi [\mnras] {10.1093/mnras/stz212}, 484, 3968

\bibitem[\protect\citeauthoryear{{Bah{\'e}}, {McCarthy}, {Balogh}  \&
  {Font}}{{Bah{\'e}} et~al.}{2013}]{bahe2013}
{Bah{\'e}} Y.~M.,  {McCarthy} I.~G.,  {Balogh} M.~L.,   {Font} A.~S.,  2013,
  \mn@doi [\mnras] {10.1093/mnras/stt109}, 430, 3017

\bibitem[\protect\citeauthoryear{{Baxter} et~al.,}{{Baxter}
  et~al.}{2017}]{baxter2017}
{Baxter} E.,  et~al., 2017, \mn@doi [\apj] {10.3847/1538-4357/aa6ff0}, 841, 18

\bibitem[\protect\citeauthoryear{{Beck} et~al.,}{{Beck}
  et~al.}{2016}]{beck2016}
{Beck} A.~M.,  et~al., 2016, \mn@doi [\mnras] {10.1093/mnras/stv2443}, 455,
  2110

\bibitem[\protect\citeauthoryear{{Boselli} \& {Gavazzi}}{{Boselli} \&
  {Gavazzi}}{2006}]{boselli2006}
{Boselli} A.,  {Gavazzi} G.,  2006, \mn@doi [\pasp] {10.1086/500691}, 118, 517

\bibitem[\protect\citeauthoryear{{Bryan} \& {Norman}}{{Bryan} \&
  {Norman}}{1998}]{bryan1998}
{Bryan} G.~L.,  {Norman} M.~L.,  1998, \mn@doi [\apj] {10.1086/305262}, 495, 80

\bibitem[\protect\citeauthoryear{{Chang} et~al.,}{{Chang}
  et~al.}{2018}]{chang2018}
{Chang} C.,  et~al., 2018, \mn@doi [\apj] {10.3847/1538-4357/aad5e7}, 864, 83

\bibitem[\protect\citeauthoryear{{Cramer}, {Kenney}, {Sun}, {Crowl}, {Yagi},
  {J{\'a}chym}, {Roediger}  \& {Waldron}}{{Cramer} et~al.}{2019}]{cramer2019}
{Cramer} W.~J.,  {Kenney} J.~D.~P.,  {Sun} M.,  {Crowl} H.,  {Yagi} M.,
  {J{\'a}chym} P.,  {Roediger} E.,   {Waldron} W.,  2019, \mn@doi [\apj]
  {10.3847/1538-4357/aaefff}, 870, 63

\bibitem[\protect\citeauthoryear{{Cui}, {Power}, {Borgani}, {Knebe}, {Lewis},
  {Murante}  \& {Poole}}{{Cui} et~al.}{2017}]{cui2017}
{Cui} W.,  {Power} C.,  {Borgani} S.,  {Knebe} A.,  {Lewis} G.~F.,  {Murante}
  G.,   {Poole} G.~B.,  2017, \mn@doi [\mnras] {10.1093/mnras/stw2567}, 464,
  2502

\bibitem[\protect\citeauthoryear{{Cui} et~al.,}{{Cui} et~al.}{2018}]{cui2018}
{Cui} W.,  et~al., 2018, \mn@doi [\mnras] {10.1093/mnras/sty2111}, 480, 2898

\bibitem[\protect\citeauthoryear{{Cybulski}, {Yun}, {Fazio}  \&
  {Gutermuth}}{{Cybulski} et~al.}{2014}]{cybulski2014}
{Cybulski} R.,  {Yun} M.~S.,  {Fazio} G.~G.,   {Gutermuth} R.~A.,  2014,
  \mn@doi [\mnras] {10.1093/mnras/stu200}, 439, 3564

\bibitem[\protect\citeauthoryear{{Diemer}}{{Diemer}}{2017}]{diemer2017a}
{Diemer} B.,  2017, \mn@doi [\apjs] {10.3847/1538-4365/aa799c}, 231, 5

\bibitem[\protect\citeauthoryear{{Diemer} \& {Kravtsov}}{{Diemer} \&
  {Kravtsov}}{2014}]{diemer2014}
{Diemer} B.,  {Kravtsov} A.~V.,  2014, \mn@doi [\apj]
  {10.1088/0004-637X/789/1/1}, 789, 1

\bibitem[\protect\citeauthoryear{{Diemer}, {Mansfield}, {Kravtsov}  \&
  {More}}{{Diemer} et~al.}{2017}]{diemer2017b}
{Diemer} B.,  {Mansfield} P.,  {Kravtsov} A.~V.,   {More} S.,  2017, \mn@doi
  [\apj] {10.3847/1538-4357/aa79ab}, 843, 140

\bibitem[\protect\citeauthoryear{{Dressler}}{{Dressler}}{1980}]{dressler1980}
{Dressler} A.,  1980, \mn@doi [\apj] {10.1086/157753}, 236, 351

\bibitem[\protect\citeauthoryear{{Fillmore} \& {Goldreich}}{{Fillmore} \&
  {Goldreich}}{1984}]{fillmore1984}
{Fillmore} J.~A.,  {Goldreich} P.,  1984, \mn@doi [\apj] {10.1086/162070}, 281,
  1

\bibitem[\protect\citeauthoryear{{Frenk} \& {White}}{{Frenk} \&
  {White}}{2012}]{frenk2012}
{Frenk} C.~S.,  {White} S.~D.~M.,  2012, \mn@doi [Annalen der Physik]
  {10.1002/andp.201200212}, 524, 507

\bibitem[\protect\citeauthoryear{{Gill}, {Knebe}  \& {Gibson}}{{Gill}
  et~al.}{2004}]{gill2004}
{Gill} S. P.~D.,  {Knebe} A.,   {Gibson} B.~K.,  2004, \mn@doi [\mnras]
  {10.1111/j.1365-2966.2004.07786.x}, 351, 399

\bibitem[\protect\citeauthoryear{{Gill}, {Knebe}  \& {Gibson}}{{Gill}
  et~al.}{2005}]{gill2005}
{Gill} S. P.~D.,  {Knebe} A.,   {Gibson} B.~K.,  2005, \mn@doi [\mnras]
  {10.1111/j.1365-2966.2004.08562.x}, 356, 1327

\bibitem[\protect\citeauthoryear{{Haines} et~al.,}{{Haines}
  et~al.}{2015}]{haines2015}
{Haines} C.~P.,  et~al., 2015, \mn@doi [\apj] {10.1088/0004-637X/806/1/101},
  806, 101

\bibitem[\protect\citeauthoryear{{Jaff{\'e}} et~al.,}{{Jaff{\'e}}
  et~al.}{2016}]{jaffe2016}
{Jaff{\'e}} Y.~L.,  et~al., 2016, \mn@doi [\mnras] {10.1093/mnras/stw984}, 461,
  1202

\bibitem[\protect\citeauthoryear{{Klypin}, {Yepes}, {Gottl{\"o}ber}, {Prada}
  \& {He{\ss}}}{{Klypin} et~al.}{2016}]{klypin2016}
{Klypin} A.,  {Yepes} G.,  {Gottl{\"o}ber} S.,  {Prada} F.,   {He{\ss}} S.,
  2016, \mn@doi [\mnras] {10.1093/mnras/stw248}, 457, 4340

\bibitem[\protect\citeauthoryear{{Knebe} et~al.,}{{Knebe}
  et~al.}{2011a}]{knebe2011a}
{Knebe} A.,  et~al., 2011a, \mn@doi [\mnras]
  {10.1111/j.1365-2966.2011.18858.x}, 415, 2293

\bibitem[\protect\citeauthoryear{{Knebe}, {Libeskind}, {Doumler}, {Yepes},
  {Gottl{\"o}ber}  \& {Hoffman}}{{Knebe} et~al.}{2011b}]{knebe2011b}
{Knebe} A.,  {Libeskind} N.~I.,  {Doumler} T.,  {Yepes} G.,  {Gottl{\"o}ber}
  S.,   {Hoffman} Y.,  2011b, \mn@doi [\mnras]
  {10.1111/j.1745-3933.2011.01119.x}, 417, L56

\bibitem[\protect\citeauthoryear{{Knollmann} \& {Knebe}}{{Knollmann} \&
  {Knebe}}{2009}]{knollmann2009}
{Knollmann} S.~R.,  {Knebe} A.,  2009, \mn@doi [\apjs]
  {10.1088/0067-0049/182/2/608}, 182, 608

\bibitem[\protect\citeauthoryear{{Larson}, {Tinsley}  \& {Caldwell}}{{Larson}
  et~al.}{1980}]{larson1980}
{Larson} R.~B.,  {Tinsley} B.~M.,   {Caldwell} C.~N.,  1980, \mn@doi [\apj]
  {10.1086/157917}, 237, 692

\bibitem[\protect\citeauthoryear{{Lin} \& {Mohr}}{{Lin} \&
  {Mohr}}{2004}]{lin2004}
{Lin} Y.-T.,  {Mohr} J.~J.,  2004, \mn@doi [\apj] {10.1086/425412}, 617, 879

\bibitem[\protect\citeauthoryear{{Moore}, {Katz}, {Lake}, {Dressler}  \&
  {Oemler}}{{Moore} et~al.}{1996}]{moore1996}
{Moore} B.,  {Katz} N.,  {Lake} G.,  {Dressler} A.,   {Oemler} A.,  1996,
  \mn@doi [\nat] {10.1038/379613a0}, 379, 613

\bibitem[\protect\citeauthoryear{{More}, {Diemer}  \& {Kravtsov}}{{More}
  et~al.}{2015}]{more2015}
{More} S.,  {Diemer} B.,   {Kravtsov} A.~V.,  2015, \mn@doi [\apj]
  {10.1088/0004-637X/810/1/36}, 810, 36

\bibitem[\protect\citeauthoryear{{More} et~al.,}{{More}
  et~al.}{2016}]{more2016}
{More} S.,  et~al., 2016, \mn@doi [\apj] {10.3847/0004-637X/825/1/39}, 825, 39

\bibitem[\protect\citeauthoryear{{Moster}, {Somerville}, {Maulbetsch}, {van den
  Bosch}, {Macci{\`o}}, {Naab}  \& {Oser}}{{Moster} et~al.}{2010}]{moster2010}
{Moster} B.~P.,  {Somerville} R.~S.,  {Maulbetsch} C.,  {van den Bosch} F.~C.,
  {Macci{\`o}} A.~V.,  {Naab} T.,   {Oser} L.,  2010, \mn@doi [\apj]
  {10.1088/0004-637X/710/2/903}, 710, 903

\bibitem[\protect\citeauthoryear{{Mostoghiu}, {Knebe}, {Cui}, {Pearce},
  {Yepes}, {Power}, {Dave}  \& {Arth}}{{Mostoghiu}
  et~al.}{2019}]{mostoghiu2019}
{Mostoghiu} R.,  {Knebe} A.,  {Cui} W.,  {Pearce} F.~R.,  {Yepes} G.,  {Power}
  C.,  {Dave} R.,   {Arth} A.,  2019, \mn@doi [\mnras] {10.1093/mnras/sty3306},
  483, 3390

\bibitem[\protect\citeauthoryear{{Patel}, {Kelson}, {Holden}, {Franx}  \&
  {Illingworth}}{{Patel} et~al.}{2011}]{patel2011}
{Patel} S.~G.,  {Kelson} D.~D.,  {Holden} B.~P.,  {Franx} M.,   {Illingworth}
  G.~D.,  2011, \mn@doi [\apj] {10.1088/0004-637X/735/1/53}, 735, 53

\bibitem[\protect\citeauthoryear{{Pe{\~n}arrubia}, {Ma}, {Walker}  \&
  {McConnachie}}{{Pe{\~n}arrubia} et~al.}{2014}]{penarrubia2014}
{Pe{\~n}arrubia} J.,  {Ma} Y.-Z.,  {Walker} M.~G.,   {McConnachie} A.,  2014,
  \mn@doi [\mnras] {10.1093/mnras/stu879}, 443, 2204

\bibitem[\protect\citeauthoryear{{Pimbblet}}{{Pimbblet}}{2011}]{pimbblet2011}
{Pimbblet} K.~A.,  2011, \mn@doi [\mnras] {10.1111/j.1365-2966.2010.17869.x},
  411, 2637

\bibitem[\protect\citeauthoryear{{Planck Collaboration} et~al.,}{{Planck
  Collaboration} et~al.}{2016}]{planck2016}
{Planck Collaboration} et~al., 2016, \mn@doi [\aap]
  {10.1051/0004-6361/201525830}, 594, A13

\bibitem[\protect\citeauthoryear{{Shin} et~al.,}{{Shin}
  et~al.}{2019}]{shin2019}
{Shin} T.,  et~al., 2019, \mn@doi [\mnras] {10.1093/mnras/stz1434}, 487, 2900

\bibitem[\protect\citeauthoryear{{Springel}}{{Springel}}{2005}]{springel2005a}
{Springel} V.,  2005, \mn@doi [\mnras] {10.1111/j.1365-2966.2005.09655.x}, 364,
  1105

\bibitem[\protect\citeauthoryear{{Springel} et~al.,}{{Springel}
  et~al.}{2005}]{springel2005b}
{Springel} V.,  et~al., 2005, \mn@doi [\nat] {10.1038/nature03597}, 435, 629

\bibitem[\protect\citeauthoryear{{Srisawat} et~al.,}{{Srisawat}
  et~al.}{2013}]{srisawat2013}
{Srisawat} C.,  et~al., 2013, \mn@doi [\mnras] {10.1093/mnras/stt1545}, 436,
  150

\bibitem[\protect\citeauthoryear{{Thomas}, {Maraston}, {Schawinski}, {Sarzi}
  \& {Silk}}{{Thomas} et~al.}{2010}]{thomas2010}
{Thomas} D.,  {Maraston} C.,  {Schawinski} K.,  {Sarzi} M.,   {Silk} J.,  2010,
  \mn@doi [\mnras] {10.1111/j.1365-2966.2010.16427.x}, 404, 1775

\bibitem[\protect\citeauthoryear{{Wen} \& {Han}}{{Wen} \&
  {Han}}{2013}]{wen2013}
{Wen} Z.~L.,  {Han} J.~L.,  2013, \mn@doi [\mnras] {10.1093/mnras/stt1581},
  436, 275

\bibitem[\protect\citeauthoryear{{White} \& {Rees}}{{White} \&
  {Rees}}{1978}]{white1978}
{White} S.~D.~M.,  {Rees} M.~J.,  1978, \mn@doi [\mnras]
  {10.1093/mnras/183.3.341}, 183, 341

\bibitem[\protect\citeauthoryear{{White}, {Cohn}  \& {Smit}}{{White}
  et~al.}{2010}]{white2010}
{White} M.,  {Cohn} J.~D.,   {Smit} R.,  2010, \mn@doi [\mnras]
  {10.1111/j.1365-2966.2010.17248.x}, 408, 1818

\bibitem[\protect\citeauthoryear{{Wu}, {Hahn}, {Wechsler}, {Behroozi}  \&
  {Mao}}{{Wu} et~al.}{2013}]{wu2013}
{Wu} H.-Y.,  {Hahn} O.,  {Wechsler} R.~H.,  {Behroozi} P.~S.,   {Mao} Y.-Y.,
  2013, \mn@doi [\apj] {10.1088/0004-637X/767/1/23}, 767, 23

\bibitem[\protect\citeauthoryear{{Zabel} et~al.,}{{Zabel}
  et~al.}{2019}]{zabel2019}
{Zabel} N.,  et~al., 2019, \mn@doi [\mnras] {10.1093/mnras/sty3234}, 483, 2251

\bibitem[\protect\citeauthoryear{{Zinger}, {Dekel}, {Kravtsov}  \&
  {Nagai}}{{Zinger} et~al.}{2018}]{zinger2018}
{Zinger} E.,  {Dekel} A.,  {Kravtsov} A.~V.,   {Nagai} D.,  2018, \mn@doi
  [\mnras] {10.1093/mnras/stx3329}, 475, 3654

\bibitem[\protect\citeauthoryear{{Z{\"u}rcher} \& {More}}{{Z{\"u}rcher} \&
  {More}}{2019}]{zurcher2019}
{Z{\"u}rcher} D.,  {More} S.,  2019, \mn@doi [\apj] {10.3847/1538-4357/ab08e8},
  874, 184

\bibitem[\protect\citeauthoryear{{van der Wel} et~al.,}{{van der Wel}
  et~al.}{2007}]{vanderwel2007}
{van der Wel} A.,  et~al., 2007, \mn@doi [\apj] {10.1086/521783}, 670, 206

\makeatother
\end{thebibliography}

%%%%%%%%%%%%%%%%%%%%%%%%%%%%%%%%%%%%%%%%%%%%%%%%%%

%%%%%%%%%%%%%%%%% APPENDICES %%%%%%%%%%%%%%%%%%%%%

%\appendix
%\section{}

%%%%%%%%%%%%%%%%%%%%%%%%%%%%%%%%%%%%%%%%%%%%%%%%%%

% Don't change these lines
\bsp	% typesetting comment
\label{lastpage}
\end{document}